\documentclass[twocolumn,showpacs,preprintnumbers,amsmath,amssymb]{revtex4}
\usepackage{graphicx}% Include figure files
\usepackage{dcolumn}% Align table columns on decimal point
\usepackage{bm}% bold math
\usepackage{amsmath}
\usepackage{amssymb}
\usepackage{float}

\renewcommand{\v}[1]{{{\bf #1}}}
\newcommand{\mat}[1]{ \overline{\bf #1} }

\renewcommand{\v}[1]{{{\bf #1}}}
\newcommand{\dyad}[1]{\mbox{$\overline{ \v{#1} }$}}

\def\dyad #1{\v{\overline{#1}}}

\def\dyad #1{\v{\overline{#1}}}
\def\dyadg #1{{\boldsymbol{\overline{#1}}}}

\def\be{\begin{equation} }
    \def\ee{\end{equation} }

\newcount\timehh\newcount\timemm
\timehh=\time \divide\timehh by 60 \timemm=\time \count255=\timehh
\multiply\count255 by -60 \advance\timemm by \count255
\def\draftheader{\slshape\today\ at
    \ifnum\timehh<10 0\fi\number\timehh\,:\,\ifnum\timemm<10 0\fi\number\timemm}%

\newcommand{\vg} [1]{\mbox{\boldmath $#1$}}

\renewcommand{\v}[1]{{{\bf #1}}}
\def\dyad #1{\v{\overline{#1}}}

\def\eeqa{\end{eqnarray}}
\renewcommand{\v}[1]{{{\bf #1}}}

\def\be{\begin{equation} }
    \def\ee{\end{equation} }

\def\Tr{\text{Tr}}

\def\PBC{periodic boundary condition}

\def\RHS{right-hand side}

\def\today{\ifcase\month\or January\or February\or March\or April\or
    May\or June\or July\or August\or September\or October\or November\or
    December \fi\space\number\day, \number\year}

\def\~{\tilde}
\def\^{\hat}

\def\I{\Im m}

\def\cal{\fam2 }
\def\Curl{\nabla\times}

\def\XXint#1#2#3{{\setbox0=\hbox{$#1{#2#3}{\int}$}
        \vcenter{\hbox{$#2#3$}}\kern-.5\wd0}}

\def\calE{\mathcal{E}_{\vg\kappa,n}}

\def\dyadcal#1{\overline{{\boldsymbol{\mathcal{#1}}}}}

\def\vcal#1{\boldsymbol{\mathcal{#1}}}

\begin{document}

\preprint{APS/123-QED}

\title{Green's Dyadic, Spectral Function, Local Density of States, and
    Fluctuation Dissipation Theorem}% Force line breaks with \\

\author{Weng Cho Chew$^1$}
\email{w-chew@illinois.edu}
\author{Wei E. I. Sha$^2$}%
\author{Qi I. Dai$^1$}%
\affiliation{$^1$Department of Electrical and Computer Engineering, University of Illinois at Urbana-Champaign, 306 N. Wright St., Urbana, IL 61801-2918\\
$^2$Department of Electrical and Electronic Engineering, University of Hong Kong, Pok Fu Lam, Hong Kong
}%

\date{\today}% It is always \today, today,

\def\VWF{vector wave function}
\def\PBC{periodic boundary condition}

\begin{abstract}
The spectral functions are studied in conjunction with the
dyadic Green's functions for various media.  The dyadic Green's
functions are found using the eigenfunction expansion method for
homogeneous, inhomogeneous, periodic, lossless, lossy, and
anisotropic media, guided by the Bloch-Floquet theorem. For the
lossless media cases, the spectral functions can be directly related
to the photon local density of states, and hence, to the
electromagnetic energy density. For the lossy case, the spectral
function can be related to the field correlation function.  Because
of these properties, one can derive properties for field
correlations and the Langevin-source correlations without resorting
to the fluctuation dissipation theorem. The results are corroborated
by the fluctuation dissipation theorem.  An expression for the local
density of states for lossy, inhomogeneous, and dispersive media has
also been suggested.
\end{abstract}

\pacs{Valid PACS appear here}% PACS, the Physics and Astronomy
\maketitle

\section{\label{sec:level1}Introduction}

The subject of thermal noise is of great interest.  Thermal noise
gives rise to Brownian motion, black body radiation, and noise in
electrical circuit known as Langevin, Johnson, or Nyquist noise
\cite{Langevin,Johnson,Nyquist,Haus}. When a system is in thermal
equilibrium with its environment, as much energy is absorbed from
the environment as is emitted to it. The fluctuation dissipation
theorem (FDT) \cite{Callen&Welton,Rytov,Kubo,Landau&Lifshitz} both
in classical and quantum form has been formulated to describe this
physical phenomenon.  Many recent works have expounded on this
phenomenon.

The recent interest in spontaneous emission, Casimir force (and the
ability to measure it), has revived the interest in this area as
well
\cite{Casimir,Casimir&Polder,Lifshitz,VanKampenEtal,Milonni,Lamoreaux,Sernelius,JohnsonSG,
    Capasso,Capasso1,RahiEtal,XiongEtal,RosaEtal}. A closely related
area is in near-field heat transfer where photons participate in
energy transfer
\cite{Narayanaswamy&Chen,GreffetEtal,JoulainEtal,SprikEtal,JonesEtal}.
Because most electromagnetic environment is dissipative, and that
quantum systems are assumed to be non-dissipative, the use of some
formulas for Casimir force calculation has given rise to
controversies:  the most well known of which is the Lifshitz formula
for Casimir force \cite{Lifshitz,VanKampenEtal,Milonni}.

The subject of thermal equilibrium also emerges in quantum transport
where electrons move from one reservoir to another when the two
reservoirs are not in equilibrium.  This gives rise to the
non-equilibrium Green's function approach.  Green's functions and
spectral functions are often defined to describe the physics of
quantum transport in micro- and nano-scopic devices
\cite{Keldysh,Datta}. Even though spectral function is often used in
quantum transport, its use in electromagnetics is less well known
\cite{Tiggelen&Kogan}.

Lossless quantum systems are easier to study.  The first
quantization of electromagnetic fields were done in lossless
systems, or weakly lossy systems
\cite{Dirac,Glauber&Lewenstein,Milonni1,Garrison&Chiao}. The theory
of quantum dissipation system has also been actively researched
\cite{Hopfield,Caldeira&Leggett,Huttner&Barnett}. These systems were
studied with a quantum system in equilibrium with a thermal bath,
sharing some commonality with FDT.
Recently, FDT has been used to
motivate the quantization of electromagnetic field in lossy media
\cite{Grunner&Welsch,DungEtal,DungEtal1,Scheel&Buhmann}.

In this paper, first, the Green's dyadics \cite{Tai,Chew}, and
spectral functions are defined and derived in terms of modal
expansions for electromagnetic system, ranging from simple
homogeneous, lossless media to anisotropic, inhomogeneous, lossy
media.  The spectral function is related to the local density of
states (LDOS) \cite{Jones&Raschke,SapienzaEtal,Busch&John} and the
density of states (DOS). Physical interpretation is given to the
electromagnetic spectral functions. Since the system is in
time-harmonic oscillation, the energy distribution of the harmonic
oscillators can be related to the generalized Planck distribution.
Finally, some equations are derived from LDOS and spectral function,
which are also derivable from the FDT.

\section{Homogeneous Unbounded Medium Dyadic Green's Function}

 The dyadic Green's function for a homogeneous medium is a solution to \cite{Tai,Chew}
 \begin{align} \label{eq:dgf_1}
 \nabla\times\nabla\times\overline{\bf G}({\bf r},{\bf r}')-k^2\overline{\bf G}({\bf r},{\bf r}')=\overline{\bf I}\delta({\bf r}-{\bf r}')
 \end{align}
 where $k^2=\omega^2\mu_0\epsilon_0$ and $\omega$ is the operating
 frequency.
 A corresponding \VWF, which is also the eigenfunction, is defined to be the solution to
 \begin{align}\label{eq:dgf_2}
 \nabla\times\nabla\times{\bf F}({\vg\kappa},{\bf r})-\kappa^2{\bf F}({\vg\kappa},{\bf r})=0
 \end{align}
 with $\kappa^2$ being the eigenvalue.  The eigenvalues of this
 system are real because the operators are Hermitian or self-adjoint
 \cite{Chew,ChewTongHu}.  For Cartesian coordinates, the \VWF s can
 be found easily. The general solution to \eqref{eq:dgf_2} is of the
 form $\v a e^{i\vg\kappa\cdot\v r}$.  In an unbounded homogeneous
 medium, $\vg\kappa$ can take on any value and is uncountably
 infinite in number. By the use of \PBC\ in a volume $V=a\times
 b\times d$, $\vg\kappa$ can be discretized and made countably
 infinite in number.  Then
 \begin{align}
 {\vg\kappa}&=\hat{x}\kappa_x+\hat{y}\kappa_y+\hat{z}\kappa_z,\,\notag\\
 \kappa_x&={2l\pi}/{a},\, \kappa_y={2m\pi}/{b},\,
 \kappa_z={2p\pi}/{d},\notag\\
  \kappa^2&=\vg\kappa\cdot\vg\kappa=(2l\pi/a)^2+(2m\pi/b)^2+(2p\pi/d)^2\notag
  \end{align}
 where $(l,m,p)$ are integer triplets.  To normalize the \VWF s, then
 \begin{align}\label{eq:dgf_3}
 {\bf M}({\vg\kappa},{\bf r})&=\frac{i{\vg\kappa}\times\hat{z}}{\kappa_s\sqrt{V}}e^{i{\vg\kappa}\cdot{\bf r}} \\\label{eq:dgf_4}
 {\bf N}({\vg\kappa},{\bf r})&=\frac{1}{\kappa}\nabla\times{\bf M}({\vg\kappa},{\bf r})=i\hat{\kappa}\times{\bf M}({\vg\kappa},{\bf r}) \\\label{eq:dgf_5}
 {\bf L}({\vg\kappa},{\bf r})&=\frac{i\hat{\kappa}}{\sqrt{V}}e^{i{\vg\kappa}\cdot{\bf r}}
 \end{align}
 where $\kappa_s=\kappa_x^2+\kappa_y^2$, and $\hat \kappa=\vg\kappa/\kappa$.
 In the above, $\v M$ is orthogonal to both $\hat z$ and $\vg\kappa$,
 $\v N$ is orthogonal to $\v M$ and $\vg\kappa$, while $\v L$ is
 longitudinal and parallel to $\vg\kappa$.     It can be shown that
 ${\bf M}$, ${\bf N}$, ${\bf L}$ above are orthonormalized, namely
 that
 \begin{align}\label{eqn:no6}
 \int_V{\bf F}_n^\dag({\vg\kappa},{\bf r})\cdot{\bf F}_{n'}({\vg\kappa}',{\bf r})\,d{\bf
    r}=\delta_{ll'}\delta_{mm'}\delta_{pp'}\delta_{nn'}
 \end{align}
 where ${\bf F}_n$, $n=1,2,3$ imply ${\bf M}$, ${\bf N}$, or ${\bf L}$.

 In the above, $\vg\kappa$ is an index associated with indices $(l,m,p)$, and that
 ${\bf M}$, ${\bf N}$, ${\bf L}$ are mutually orthonormal. Hence, assuming the completeness of these eigenfunctions,
 then \footnote{Usually, in the physics notation, for inner
    products, the transpose of
    the left vector is not explicit as in \eqref{eqn:no6}, but
    in linear algebraic notation, it is explicit as is used here.
    For outer products here, one will explicitly
    put
    the transpose of the right vector as well.}
 \begin{align}
 \overline{\bf I}\delta({\bf r}-&{\bf r}')=\nonumber\\
 &\sum_{\vg\kappa}\left[{\bf M}({\vg\kappa},{\bf r}){\bf a}^t({\vg\kappa})+{\bf N}({\vg\kappa},{\bf r}){\bf b}^t({\vg\kappa})+
 {\bf L}({\vg\kappa},{\bf r}){\bf c}^t({\vg\kappa})\right]
 \end{align}
 Using orthonormality, it follows that \footnote{The completeness of these eigenfunctions can be proved
    along similar line as that for Fourier series.}
 %\begin{widetext}
 \begin{align}
 \overline{\bf I}\delta({\bf r}-{\bf r}')=&
 \sum_{\vg\kappa}\left[{\bf M}({\vg\kappa},{\bf r}){\bf M}^\dag
 ({\vg\kappa},\v r')+\right.\notag\\&\left.
{\bf N}({\vg\kappa},{\bf r}){\bf N}^\dag ({\vg\kappa},\v r')  +
 {\bf L}({\vg\kappa},{\bf r}){\bf L}^\dag({\vg\kappa},\v r')\right]
 \end{align}
%\end{widetext}
 or more concisely,
 \begin{align}
 \overline{\bf I}\delta({\bf r}-{\bf r}')=\sum_{{\vg\kappa},n}{\bf F}_n({\vg\kappa},{\bf r}){\bf F}_n^\dag({\vg\kappa},{\bf r}')
 \end{align}
 where ${\bf F}_n$, $n=1,2,3$ denote ${\bf M}$, ${\bf N}$, or ${\bf L}$, respectively.   If $\hat z$ is assumed to be vertical, one can think of
 $\v M$ as horizontal, $\v N$ as vertical (with respect to $\v M$ and $\vg\kappa$), and $\v L$ as longitudinal.  By letting $\dyad G(\v r,\v
 r')$
 in
 \eqref{eq:dgf_1} as
 \begin{align}
 \overline{\bf G}({\bf r},{\bf r}')=\sum_{{\vg\kappa},n}{\bf F}_n({\vg\kappa},{\bf r}){\bf f}^t_n({\vg\kappa})
 \end{align}
 it follows that
 \begin{align}\label{eq:dgf_10}
 \overline{\bf G}({\bf r},{\bf r}',k)=\sum_{{\vg\kappa},n}\frac{{\bf F}_n({\vg\kappa},{\bf r}){\bf F}_n^\dag({\vg\kappa},{\bf r}')}{\kappa_n^2-k^2}
 \end{align}
 where $\kappa_n^2=\vg\kappa\cdot\vg\kappa$ when $n=1,2$, and $\kappa_3=0$.  Here, $n=1,2$ are used
 to index the transverse modes $\v M$ and $\v N$, while $n=3$
 indexes the longitudinal mode $\v L$.  The above is undefined
 unless a small infinitesimal loss is associated with $k$.

 The above is very similar to the solution of a modal cavity
 driven by a time-harmonic source with operating frequency $\omega$
 where $k=\omega/c$.  The $\v M$ and $\v N$ modes are
 dynamic modes with resonant wave number $\kappa_n$, while the $\v
 L$ modes are static modes with zero resonant wave number.
 But in this case, the modes are generated using \PBC\ rather
 than a cavity wall. In the above, \eqref{eq:dgf_3} to
 \eqref{eq:dgf_5} represent travelling waves while if cavity
 modes were used, they would represent standing waves.

 \section{The Density of States}

 The density of states is the number of states (DOS) or modes per
 unit energy interval.  For each eigenmode, there corresponds a
 frequency $\omega=\kappa c$.
 Since photons are described here, the relation between photon energy
 and wavenumber is $\mathcal{E}_{\vg\kappa} =\hbar
 \omega_{\vg\kappa}= \hbar \kappa c$, or $\kappa =
 \mathcal{E}_{\vg\kappa}/(\hbar c)$,
 $\hbar$ is the Planck constant, and $c$ is the velocity of light.

 Since
 \begin{equation}{\label{eq:dos_11}}
    {\vg\kappa} = {\hat x}\frac{2l\pi }{a} + {\hat y}\frac{2m\pi}{b} +
    {\hat z}\frac{2p \pi}{d}
 \end{equation}
 each mode occupies a volume of ${(2\pi)^{3}}/{(abd)} =
 {8\pi^{3}}/{V}$ in ${\vg\kappa}$-space. The number of modes or
 states in a spherical shell between $\kappa$ and $\kappa+\Delta
 \kappa$ is
 \begin{equation}{\label{eq:dos_12}}
    \Delta N = 2 \times 4\pi \kappa^{2} \Delta \kappa/(8\pi^{3}/V) =
    \frac{1}{\pi^{2}} \kappa^{2}\Delta \kappa V
 \end{equation}
 where one assumes that there are two polarizations for every mode.
 Notice that if $\Delta N$ is normalized by $V$, it is independent of
 $V$.

 A succinct way to write the density of states is
 \begin{equation}{\label{eq:dos_13}}
    D(E) = \sum_{\vg\kappa,n} \delta (E - \mathcal{E}_{\vg\kappa,n})
 \end{equation}
 where $\mathcal{E}_{\vg\kappa,n} = \hbar \kappa c$, $n=1,2$
 represent the two polarizations, horizontal and vertical, and
 $E=\hbar \omega$. Here, $\kappa^2=|\vg\kappa|^2$ is the eigenvalue
 of \eqref{eq:dgf_2}, which are non-zero for $\v M$ and $\v N$ modes.
 Since $\vg\kappa$ is associated with indices $(l,m,p)$, there are
 many degeneracies in the eigenvalue $\kappa^2$ in \eqref{eq:dgf_2}:
 all $\vg\kappa$ vectors from the origin to the sphere defined by
 $|\vg\kappa|=\kappa$ are degenerate. Also, in the above, one only
 retains states or modes with non-zero resonant wavenumbers. The $\v
 L$ modes are not counted in the above since they have zero resonant
 wavenumber in \eqref{eq:dgf_2}.

 Equation (\ref{eq:dos_13}) is like a spectroscopy function showing
 spikes whenever $E$ coincides with one of the eigenstates with
 energy $\mathcal{E}_{\vg\kappa,n}$. The number of states in an
 energy interval $E$ and $E + \Delta E$ is given by
 \begin{equation}\label{eq:dos_14}
    \int_{E}^{E + \Delta E} D(E) dE = \Delta N, \quad
    ({E<\mathcal{E}_{\vg\kappa,n}<E +\Delta E})
 \end{equation}
 It is seen that \eqref{eq:dos_13} is in fact the density of states
 per unit energy interval because integrating it over an energy
 interval gives the number of states in that interval, as shall be
 shown below.

 In the limit when $V\rightarrow\infty$, one can convert the
 summation in \eqref{eq:dos_13} into an integral by the substitution
 that $\sum_{\vg\kappa}=V/(2\pi)^3 \sum \Delta \kappa_x\Delta
 \kappa_y \Delta \kappa_z=V/(2\pi)^3 \int d\vg\kappa$. Namely,
 \begin{equation}\label{eq:dos_15}
 \begin{aligned}
    D(E) &= 2\times \frac{V}{(2\pi)^3} \int d \vg\kappa\delta(E-\hbar
    \kappa c) \\
    & =\frac{V}{4\pi^3}\int_0^\infty 4\pi \kappa^2 d\kappa
    \delta(E-\hbar \kappa c)=\frac{Vk^3}{\pi^2 E}
    \end{aligned}
 \end{equation}
 where it is noted again here that $E=\hbar k c=\hbar \omega$. In the
 above, $\mathcal{E}_{\vg\kappa,n}=\hbar \kappa c$ has no angular
 dependence, and hence, $\int d\vg\kappa=\int_0^\infty 4\pi
 \kappa^2d\kappa$.
 Consequently
 \begin{align}\label{eq:dos_16}
    \Delta N = D(E)\Delta E = \frac{Vk^2}{\pi^2}\Delta k
 \end{align}
 which is the same as \eqref{eq:dos_12} except that one replaces
 $\kappa$ in \eqref{eq:dos_12} with $k$ here since $E=\hbar \omega$
 is the counter.  The above is self consistent with
 \eqref{eq:dos_13}.

 \section{Spectral Function and Local Density of States}\label{sect4}

 The photon (electromagnetics) local density of states (LDOS) is
 defined as
 \begin{equation}\label{eq:ldos_15}
    D({\v r},E) = \sum_{\vg\kappa,n} |{\v F}_n({\vg\kappa},{\v
        r})|^{2}\delta(E - \mathcal{E}_{\vg\kappa,n})
 \end{equation}
 where $n=1,2$ represent the two polarizations, and $\vg\kappa$
 indicates the plane wave direction of the eigenfunction since they
 are traveling plane waves. The above has the property that
 \begin{equation}\label{eq:ldos_16}
    D(E) = \int_{V}d{\v r} D({\v r},E)
 \end{equation}
 since
 \begin{equation}\label{eq:ldos_17}
    \int_{V}|{\v F}_n({\vg\kappa},{\v r})|^{2}d{\v r} = 1
 \end{equation}
 The above implies that the LDOS, when integrated over $\v r$, yields
 DOS.  Therefore, LDOS has the unit of states per unit energy per
 unit volume.   It also associates different probabilities of finding
 a photon in different locations denote by $\v r$.

 %%%%%%%%%%%%%%%%%%%%%%%%%%%%%%%%%%%%%%%%%%%%%%%%%%%%%%%%%%%%%%%%%%%%%%%%%%%%%%%%%%%%%%%%%%%%%%%%%%%%%%%%%%%%%%%%%%%%%%%%%%%%%%%%%%%%%%%%%%%%%%%%%%%%
 %%%%%%%%%%%%%%%%%%%%%%%%%%%%%%%%%%%%%%%%%%%%%%%%%%%%%%%%%%%%%%%%%%%%%%%%%%%%%%%%%%%%%%%%%%%%%%%%%%%%%%%%%%%%%%%%%%%%%%%%%%%%%%%%%%%%%%%%%%%%%%%%%%%%
 %%%%%%%%%%%%%%%%%%%%%%%%%%%%%%%%%%%%%%%%%%%%%%%%%%%%%%%%%%%%%%%%%%%%%%%%%%%%%%%%%%%%%%%%%%%%%%%%%%%%%%%%%%%%%%%%%%%%%%%%%%%%%%%%%%%%%%%%%%%%%%%%%%%%

 The LDOS can be related to the dyadic Green's function as shall be
 shown next.  To this end, a spectral function is defined as
 %\footnote{In the literature, this is simply known as the spectral
 %function.}
 \begin{align}\label{eq:ldos_18}
    \dyad A(\v r,\v r')=i\left[\dyad G(\v r, \v r')-\dyad G^\dagger (\v
    r', \v r)\right]
 \end{align}
 The second Green's operator is the adjoint of the first one. The
 Green's dyadic is an integral operator acting on a functional
 Hilbert space made of 3 vectors.  The adjoint of this Green's dyadic
 operator is defined to satisfy the following relationship
 \begin{align}\label{eqn22}
    \int d\v r \v f_2^\dag(\v r)\cdot \int & d\v r' \dyad G(\v r, \v
    r')\cdot \v f_1(\v r') =\nonumber\\
    & \int d\v r' \int d\v r \left[ \dyad G^a(\v
    r',\v r)\cdot \v f_2(\v r)\right]^\dag\cdot \v f_1(\v r')
 \end{align}
 In this manner, the transpose of a scalar integral operator is such
 that $g^t(\v r,\v r')=g(\v r',\v r)$.  This is similar to that for
 matrices, where matrix elements of transpose matrices are related by
 $\left[ \dyad A^t\right]_{ij}=\left[ \dyad A \right]_{ji}$. Hence,
 one swaps the argument of the second Green's dyadic in
 \eqref{eq:ldos_18} to make it the adjoint of the first Green's
 dyadic operator. To elaborate, from the above, one obtains
 \begin{equation}
    \dyad G^{a}(\v r',\v r)=\dyad G^\dag(\v r, \v r'),{\rm{\quad or \quad
    }} \dyad G^{a}(\v r,\v r')=\dyad G^\dag(\v r', \v r)
 \end{equation}
 In this manner, the spectral function so defined above is a
 self-adjoint operator.

 It can be shown that the above represents a Hermitian system. For
 a reciprocal Green's function, it further has the property that
 $\dyad G^t(\v r,\v r')=\dyad G(\v r',\v r)$.  Hence, the spectral
 function can be written as
 \begin{align}\label{eq:ldos_19}
    \dyad A(\v r,\v r')=i\left[\dyad G(\v r, \v r')-\dyad G^* (\v r, \v
    r')\right]=-2\Im m \left[\dyad G(\v r, \v r')\right]
 \end{align}
 From \eqref{eq:dgf_10}, it can shown that
 \begin{align}\label{eq:ldos_20}
    \overline{{\bf G}}^*({\bf r}, {\bf r}')&=\sum_{{\vg\kappa},n}
    \frac{{\bf F}^\ast_n({\vg\kappa},{\bf r}){\bf
            F}_n^t({\vg\kappa},{\bf r}')}{\kappa_n^2-k^{*2}}\nonumber\\& =
    \sum_{{\vg\kappa},n} \frac{{\bf F}_n({\vg\kappa},{\bf r}){\bf
            F}^\dag_n({\vg\kappa},{\bf r}')}{\kappa_n^2-k^{*2}}
 \end{align}
 where $\kappa_n^2=\vg\kappa\cdot \vg\kappa$, and the last equality
 is obtained by letting $ {\vg\kappa}\rightarrow -{\vg\kappa} $ and
 one notices that $ {\bf F}_n({-\vg\kappa},{\bf r}) = {\bf
    F}^*_n({\vg\kappa},{\bf r})$. Moreover, it also follows from that
 the eigenvalues of \eqref{eq:dgf_2} is real and hence, $\kappa_n^2$
 is real.
 Therefore
 \begin{equation}
 \begin{aligned}\label{eq:ldos_21}
    \dyad A(\v r,\v r') &=-2\Im m\left[\overline{{\bf G}}({\bf r}, {\bf
        r}')\right]=i\left[\overline{{\bf G}}({\bf r}, {\bf
        r}')-\overline{{\bf G}}^*({\bf r}, {\bf r}')\right] \\
    &=
    -2\sum_{{\vg\kappa},n}{\bf F}_n({\vg\kappa},{\bf r}){\bf
        F}^\dag_n({\vg\kappa},{\bf r}')\Im
    m\left(\frac{1}{\kappa_n^2-k^{2}}\right)
 \end{aligned}
 \end{equation}
 One can rewrite
 \begin{align}\label{eq:ldos_22}
    \frac{1}{\kappa_n^2-k^2}=\frac{1}{2\kappa_n}\left(\frac{1}{\kappa_n-k}+\frac{1}{\kappa_n+k}\right)
 \end{align}
 If $ k\rightarrow k +i\eta $ to have a small imaginary part to mean
 loss, then
 \begin{equation}
 \begin{aligned}\label{eq:ldos_23}
    \Im m \left(\frac{1}{\kappa_n-k-i\eta}\right) &= \Im m
    \left[\frac{i\eta}{(\kappa_n-k)^2+\eta^2}\right] \\
    &=
    \frac{\eta}{(\kappa_n-k)^2+\eta^2}
 \end{aligned}
  \end{equation}
 It can be shown that
 \begin{align}\label{eq:ldos_24}
    \lim_{\eta\rightarrow 0} \frac{\eta}{x^2+\eta^2} = \pi \delta(x)
 \end{align}
 Comparing, and using the above in \eqref{eq:ldos_21}, it means that
 %\begin{widetext}
 \begin{align}\label{eq:ldos_25}
    \dyad A(\v r, \v r')&= -2\Im m\left[\overline{{\bf G}}({\bf r}, {\bf
        r}')\right] = -2\sum_{{\vg\kappa},n}{\bf F}_n({\vg\kappa},{\bf
        r}){\bf F}^\dag_n({\vg\kappa},{\bf
        r}')\notag\\&\quad\quad\quad\cdot\frac{\pi}{2\kappa_n}\left[\delta(\kappa_n-k)+\delta(\kappa_n+k)\right]
 \end{align}
 % \end{widetext}
 %
 Furthermore, $ \delta(ax) = a^{-1}\delta(x)$, $
 \kappa_n=\mathcal{E}_{{\vg\kappa},n}/{\hbar c} $, and $k = E/{\hbar
    c}$, imply that
 \begin{align}\label{eq:ldos_26}
    \Im m\left(\frac{1}{\kappa_n^2-k^{2}}\right)
    = \frac{\pi E}{2k^2}\left[\delta(E-\mathcal{E} _{\vg\kappa,n}) -
    \delta(E + \mathcal{E}_{\vg\kappa,n})\right]
 \end{align}
 Consequently,
 \begin{equation}
 \begin{aligned}\label{eq:ldos_27}
    \dyad A(\v r,\v r') & =-2\Im m\left[\overline{{\bf G}}({\bf r}, {\bf
        r}')\right]\\
    & = -\frac{\pi E}{k^2}\sum_{{\vg\kappa},n} {\bf
        F}_n({\vg\kappa},{\bf r}){\bf F}^\dag_n({\vg\kappa},{\bf
        r}')\delta(E-\mathcal{E}_{{\vg\kappa},n})
 \end{aligned}
 \end{equation}
 Here, $ \delta(E + \mathcal{E}_{{\vg\kappa},n}) $ is ignored since $
 E + \mathcal{E}_{{\vg\kappa},n} \neq 0 $, and
 $\mathcal{E}_{{\vg\kappa},n}>0$ for all $\vg\kappa$. Another point
 to note regards the longitudinal modes ($n=3$) with zero resonant
 frequency $\mathcal{E}_{{\vg\kappa},n} = 0$. If one keeps track of
 modes only with $\mathcal{E}_{{\vg\kappa},n}>0$, they need not be
 included in the above expression.

 If $ {\bf r}= {\bf r}' $, and noticing that by taking the trace, one
 has $\Tr\left[{\bf F}_n({\vg\kappa},{\bf r}){\bf
    F}^\dag_n({\vg\kappa},{\bf r})\right] = \sum\limits_{i=1}^{3}
 F_{ni}({\vg\kappa},{\bf r})F_{ni}^\ast({\vg\kappa},{\bf r}) = |{\bf
    F}_n({\vg\kappa},{\bf r})|^2$, it means that
 \begin{align}\label{eq:ldos_28}
    D(\v r, E)= -\frac{k^2}{\pi E}\Tr\left[\dyad A(\v r,\v r)\right]
    =\frac{2k^2}{\pi E}\Tr \left\{ \Im m\left[\overline{{\bf G}}({\bf
        r},{\bf r})\right]\right\}
 \end{align}
 where $D(\v r,E)$ is the local density of states as given by
 \eqref{eq:ldos_15}. Hence, the LDOS can be found once the imaginary
 part of the dyadic Green's function is known.  In the above, the
 spectral function and the Green's dyadic are implicit functions of
 $\omega$ and hence, $E$.

 \subsection{Photon Energy Density}

 A photon with frequency $\omega$ has quantization energy given by
 $E=\hbar \omega$. It is the result of energy quantization of the
 quantum harmonic oscillator.   If the quantum harmonic oscillator is
 in thermal equilibrium with another system, it can be shown that the
 average photon number in the photon number state of the oscillator
 is \cite{Gerry&Knight,Fox}
 \begin{equation}
    \bar n=\frac{1}{e^{\hbar\omega/(k_BT)}-1}
 \end{equation}
 Then the average energy of the photon state is given by
 \begin{align}\label{eq:ped_29}
    \Theta({E=\hbar
        \omega})=\left(\bar n+\frac{1}{2}\right) \hbar \omega=
    \frac{1}{2}\hbar\omega+\frac{\hbar\omega}{e^{\hbar\omega/(k_BT)}-1}
 \end{align}
 The above result can be derived from the quantum statistical mechanics of
 a quantum
 harmonic oscillator.
 In the limit when $\hbar\rightarrow 0$, the above becomes the
 classical result showing that $\Theta(E)=k_B T$: this is in agreement with the
 equipartition of energy theorem with $0.5 k_BT$ per degree of freedom. A harmonic oscillator has two degrees of freedom,
 one for storing kinetic energy while another one for storing potential energy \cite{Feynman}.

 If $T\rightarrow 0$, then
 $\Theta(E)=(1/2)\hbar\omega$, which is also a result from quantum
 mechanics.  It implies that the quantum harmonic oscillator has nonzero energy even if it is in the
 ground state: this is called the zero point energy.  This zero point
 energy is there even when $\bar n=0$, or with no photon
 \cite{Gerry&Knight,Fox} is present on average.  This is the source
 of vacuum fluctuation of the electromagnetic field, and is the
 source of the Casimir force at zero temperature \cite{Milonni}.

 Hence, if the density of states of the system is known, the average
 photon energy of the system is given by
 \begin{align}\label{eq:ped_30}
    \langle {\cal E}\rangle =\int_{0}^{\infty} D(E) \Theta(E) dE
 \end{align}
 In the above, $D(E)\Theta(E)dE$ is the average photon energy in the
 energy interval $dE$, and hence, when integrated yields the average
 energy of the system. If one lets $E=\hbar \omega$, and writes
 \begin{equation}
    D(E)\Theta(E)dE=D(E)\Theta(E)\hbar d\omega
 \end{equation}
 then one can identify, after using \eqref{eq:dos_15} for $D(E)$ that
 \begin{equation}
    U(\omega)=\frac{k^3}{\pi^2E}\Theta(E)\hbar=\frac{\omega^2}{\pi^2c^3}\left(\frac{1}{2}\hbar\omega+\frac{\hbar\omega}{e^{\hbar\omega/(k_BT)}-1}\right)
 \end{equation}
 is just the black body energy per unit volume per unit radian
 frequency according to the generalized Planck's radiation law,
 except that the zero-point energy is absent in the original Planck's
 law.

 If the local density of states is known, then the average photon
 energy per unit volume or the photon energy density  at the location
 $\v r$ is given by
 \begin{align}\label{eq:ped_31}
    \langle {\cal E}(\v r)\rangle =\int_{0}^{\infty} D(\v r,E) \Theta(E)
    dE
 \end{align}
 Given that $D(\v r,E) \Theta(E)$ implies energy per unit energy per
 unit volume, it is prudent to define a field correlation
 function \footnote{This is due to hindsight that will be confirmed in
    Sect. 8.}
 \begin{equation}\label{eqn39}
 \begin{aligned}
    \dyadcal C(\v r,\v r', \omega) &=\langle \vcal E(\v r, \omega)\vcal
    E^\dag(\v r', \omega)\rangle=-\frac{\omega\mu_0}{\pi} \dyad A (\v
    r,\v r',\omega)\Theta(\hbar\omega) \\
    & =\frac{2\omega\mu_0}{\pi}\Im m
    \left[ \dyad G (\v r,\v r',\omega)\right]\Theta(\hbar\omega)
    \end{aligned}
 \end{equation}
 where $E=\hbar\omega$. In this manner, the LDOS can be directly
 related to this field correlation function as
 \begin{align}\label{eqn40}
    D(\v r,E)\Theta(E)dE=D(\v r,E)\Theta(E)\hbar d\omega=\Tr [\epsilon_0
    \dyadcal C(\v r, \v r, \omega)]d\omega
 \end{align}
 In the above, $\dyadcal C$ has the unit of field square per radian
 frequency.
 The factor of 1/2 is not needed here because the above represents
 the energy density in both the electric field and the magnetic
 field.   For a homogeneous medium, they are equal to each other.
 With the above, we can define a density of states per unit radian
 frequency, $\tilde D(\v r,\omega)$, so that
 \begin{align}
    \tilde D(\v r,\omega)\Theta(\hbar\omega)d\omega=D(\v r,E)\Theta(E)
    dE=D(\v r,E)\Theta(E)\hbar  d\omega
 \end{align}
 or
 \begin{align}
    \tilde D(\v r,\omega)=D(\v r,E)\hbar=\frac{2\omega\mu_0}{\pi}\Tr\Im
    m \left[ \dyad G (\v r,\v r,\omega)\right]
 \end{align}
 The relation expounded in \eqref{eqn39} can also be confirmed in
 Section \ref{sect8}, and are commensurate with FDT as shall be shown
 later.

 \subsection {Physical Interpretation--Lossless Case}

 The spectral function is formed by the difference of two Green's
 function: it describes the response of a point source in equilibrium
 with its environment.  If the Green's dyadic describes a point
 source field that is leaving the source point, the adjoint of the
 Green's dyadic describes a field that is returning back to the point
 source.  If the Green's function satisfies causality in the time
 domain, it is also called the retarded Green's function, or the
 forward Green's function.  The adjoint Green's function is
 non-causal, and it is also called the backward Green's fuction, the
 advanced Green's function, or a time reversed Green's function in
 the time domain.  A forward Green's function fore propagates the
 field from a source point to a field point, while a backward Green's
 function back propagates the field from the field point to the
 source point.
 For instance, in the homogeneous medium case, one Green's function
 yields an outgoing field from the source point, while the other one
 absorbs an incoming field to the source point.  The reactive fields
 \cite{Chew} that represent stored energy in these two Green's
 functions cancel each other, and only the radiative fields remain in
 the difference.

 In the lossless medium, when the frequency is off resonance,
 $\kappa_n\ne k$, the forward Green's function and the reversed
 Green's function cancel each other exactly, but when $\kappa_n=k$,
 they do not cancel each other giving rise to spikes in the spectral
 function.
 It yields the density of states of the system in \eqref{eq:dos_13}
 and the local density of states in \eqref{eq:ldos_15}. When the
 spectral function is multiplied by $\Theta(E)$, the average photon
 energy per state, it yields a quantity proportional to the energy
 density.
 Therefore, we have suggestively written the spectral function in
 \eqref{eqn39} so that it is related to the correlation function of
 the field at two points.  This relationship also follows from
 Section \ref{sect8} and also the fluctuation dissipation theorem.
 Hence, the spectral function motivates a result that can be drawn
 from the fluctuation dissipation theorem.

 \section{Line Broadening and Loss}

 Notice that \eqref{eq:dos_13} resembles a set of spectroscopic lines
 that are infinitely sharp because there is no loss in the system
 corresponding to infinite Q. If loss is introduced, such
 spectroscopic lines will be broadened due to finite Q. To see this,
 the delta function in \eqref{eq:ldos_23} will be broadened if the
 loss is not infinitesimally small.

 It is to be noted that \eqref{eq:dgf_10} is still valid even when
 $k$ corresponds to a lossy medium or is a complex number.
 The generalized spectral function then becomes
 \begin{equation}
 \begin{aligned}\label{eq:lbal_38}
    \dyad A(\v r,\v r')&=-2\Im m\left[\overline{{\bf G}}({\bf r}, {\bf
        r}')\right]=i\left[\overline{{\bf G}}({\bf r}, {\bf
        r}')-\overline{{\bf G}}^*({\bf r}, {\bf r}')\right]\\
    &=
    -2\sum_{{\vg\kappa},n}{\bf F}_n({\vg\kappa},{\bf r}){\bf
        F}^\dag_n({\vg\kappa},{\bf r}')\gamma(\kappa_n,k)
 \end{aligned}
 \end{equation}
 where
 \begin{equation}
 \begin{aligned}\label{eq:lbal_39}
    \gamma(\kappa_n,k) &= \Im m\left(\frac{1}{\kappa_n^2-k^{2}}\right) \\
    &=\Im
    m
    \left[\frac{1}{2\kappa_n}\left(\frac{1}{\kappa_n-k}+\frac{1}{\kappa_n+k}\right)\right]
 \end{aligned}
 \end{equation}
 It can be shown that in the limit when $V\rightarrow\infty$, the
 summation becomes an integral, giving \footnote{Derivation available
    upon request to the authors.}
 \begin{align}\label{eq:lbal_40}
    \sum_{{\vg\kappa},n} \gamma(\kappa_n,k)=2\times \frac{V}{8\pi^3}
    \int d\vg\kappa \gamma(\kappa,k)=\frac{V}{2\pi} k'
 \end{align}
 where $k=k'+i k''$, $k'$ and $k''$ are real numbers, but the
 integral is independent of $k''$, the loss. The multiplication by 2
 is necessary because there are two polarizations per mode. In the
 above, $k=\omega\sqrt{\mu_0\epsilon_0}$ where for a lossy medium,
 $\mu_0$ and $\epsilon_0$ are complex numbers.  But by the
 Kramers-Kronig relation \cite{Kramers&Kronig}, they have to be
 functions of frequency $\omega$: their real and imaginary parts are
 related by Hilbert transforms \cite{Chew}.

 A generalized local density of states can be defined as
 \begin{equation}\label{eq:lbal_15}
    D({\v r},E) = \sum_{\vg\kappa,n} |{\v F}_n({\vg\kappa},{\v
        r})|^{2}\Gamma(\mathcal{E}_{\vg\kappa,n},E)
 \end{equation}
 where
 $$\Gamma(\mathcal{E}_{\vg\kappa,n},E)=\frac{2k'^2}{\pi E}\gamma(\kappa_n,k)
 $$
 with $\kappa_n=\mathcal{E}_{\vg\kappa,n}/(\hbar c)$, $E=\hbar
 \omega$, and $c$ is the velocity of light in vacuum. In the above,
 $\kappa_n$ as defined in \eqref{eq:dgf_10} remains real when $k$ is
 complex due to material loss.
 In this manner,
 $$\lim_{ k''\rightarrow 0} \Gamma(\mathcal{E}_{\vg\kappa,n},E)=\delta
 (E-\mathcal{E}_{\vg\kappa,n})$$ Then a generalized density of states
 becomes
 \begin{equation}\label{eq:lbal_16}
    D(E) = \sum_{\vg\kappa,n} \Gamma(\mathcal{E}_{\vg\kappa,n},E)
 \end{equation}
 The above has a nice physical meaning since it can be shown that
 \begin{equation}
 \begin{aligned}
    D(E)&=\sum_{\vg\kappa,n} \Gamma(\mathcal{E}_{\vg\kappa,n},E)=
    \frac{2k'^2}{\pi E}\sum_{\vg\kappa,n} \gamma(\kappa_n,k)\\
    &= 2\times
    \frac{V}{8\pi^3} \frac{2k'^2}{\pi E} \int_{-\infty}^{\infty}
    d\vg\kappa \gamma(\kappa,k)=\frac{k'^3V}{\pi^2 E}
 \end{aligned}
 \end{equation}
 Therefore, the spectral function can still be related to the LDOS in
 the lossy case.  The DOS in this case is related to $k'$, the real
 part of the $k$ where the $\gamma(\kappa,k)$ function peaks on the
 real axis of $\kappa$.  If we choose $V=1$, then the DOS is in terms
 of the number of states per unit volume per unit energy.  We will
 show in Section \ref{sect8} that this relationship exists even for a
 general lossy, anisotropic, inhomogeneous medium.

 \subsection{Physical Interpretation--Lossy Case}

 The above formula is similar to \eqref{eq:dos_15} excepted that $k$
 is now replaced by $k'$, the location on the complex $\kappa$ plane
 where $\gamma(\kappa,k)$ peaks. The $\Gamma$ function, which is the
 normalized $\gamma$ function, behaves like a broadened delta
 function, and in the limit of no loss case, reverts back to a delta
 function.  In the lossless case, each $k$ is associated with modes
 with $\kappa=k$ as in \eqref{eq:dos_13} and \eqref{eq:ldos_15}, but
 in the lossy case, there is a cluster of modes with  $\kappa$ values
 associated with the peak of the $\gamma$ function when $\kappa=k'$
 as indicated in \eqref{eq:lbal_39} and \eqref{eq:lbal_40}.

 In a lossy medium, the first term of the spectral function describes
 a decaying field, but the second term, a back propagating field,
 describes a growing field.  This can be thought of as a lossy system
 in equilibrium with a thermal bath.  The loss in the system is
 accompanied by Langevin sources \cite{Haus} induced by the thermal
 excitation of the environment.  These sources produce a field that
 grows, instead of decays with distance.

 In either cases, the spectral function describes a system in thermal
 equilibrium.  In the lossless case, the system is in thermal
 equilibrium with sources at infinity, while in the lossy case, the
 system is in equilibrium with the Langevin sources due to the loss
 in the medium. In both cases, the system is in time harmonic motion
 because of thermal equilibrium: as much energy is supplied to the
 system as it is lost, so that the harmonic oscillators in the
 vacuum, atoms, and molecules are in simple time-harmonic
 oscillation.  (Notice that the spectral function is Hermitian
 because it describes an energy conserving system.)

 %In the lossless case, when the system is excited to be in a simple
 %time harmonic motion with frequency $\omega$, only one mode or its
 %degenerate modes are excited to be in time harmonic motion as
 %indicated in \eqref{eq:dos_13} and \eqref{eq:ldos_15}.

 But in the lossy case, when the system is excited to be in a simple
 time-harmonic motion, a cluster of modes are excited in unison to be
 in time-harmonic motion.  When these harmonic oscillators are
 quantized to be associated with a quantum or packet of energy
 $\hbar\omega$, the energy is distributed over the cluster of modes
 with different $\mathcal{E}_{\vg\kappa,n}$, and weighted with
 different coefficients according to $\gamma$.  These modes can be
 given a probabilistic interpretation as in quantum mechanics.
 Because of the above physical interpretation, one can use
 \eqref{eq:ped_30} to \eqref{eq:ped_31} to obtain the photon energy
 and photon energy density of a lossy system as well.

 \section{The Periodic Structure Case}

 The periodic structures can be assumed to be perfect electric
 conductors (PEC) \footnote{The perfect magnetic conductor (PMC) case
    can be similarly treated.}. In this case, the electric \VWF s or
 eigenfunctions are defined to be the solutions to eigenvalue problem
 \begin{align}\label{eq:psc_36}
    \nabla \times \nabla \times \v{F}_{e,n} (\vg\kappa, \v{r}) -
    \kappa_n^2 \v{F}_{e,n}(\vg\kappa, \v{r)} = 0
 \end{align}
 where $\vg \kappa$ is the same as defined as above \eqref{eq:dgf_3}.
 By choosing a periodic PEC (or PMC) structure, the above eigenvalue
 problem is similar to \eqref{eq:dgf_2} except for the boundary
 condition to be satisfied on the surface of the PEC (or PMC).  The
 above problem is easily proved to be Hermitian or self-adjoint, and
 hence, the eigenvalues are real.
 The above eigenvalue problem, in accordance with Bloch-Floquet
 theorem, will yield eigen-solutions where \cite{Busch&John}
 \begin{align}\label{eq:psc_37}
    \v{F}_{e,n}(\vg\kappa,\v{r}) = \sum_{\v{G}} \sum_{\lambda}
    F_{\v{G}}^\lambda \v{e}_{\v{G}}^\lambda e^{i (\vg\kappa +
        \v{G})\cdot \v{r}}
 \end{align}
 and $\v{G}$'s are the reciprocal lattice vectors, and
 $\lambda\in\{1,2,3\}$, with $\v{e}_{\v{G}}^1$, $\v{e}_{\v{G}}^2$,
 $\v{e}_{\v{G}}^3$ forming an orthogonal triad unit vectors, and
 $\v{e}_{\v{G}}^3$ is parallel to $\vg\kappa+\v G$. Hence,
 $\lambda\in\{1,2\}$ correspond to transverse modes, while
 $\lambda=3$ corresponds to the longitudinal mode.

 The above is solved with the boundary condition that $\hat{n} \times
 \v{F}_{e,n} = 0$ on the PEC surface of the scatterer where $\hat n$
 is the unit surface normal. In this case, the eigenfunctions and
 eigenvalues have to be found numerically. It is to be noted that
 $\kappa_n^2(\vg\kappa)$ is the eigenvalue of the equation
 \eqref{eq:psc_36} with different choices of the Bloch-Floquet wave
 vector $\vg\kappa$.  Hence, it is a function of $\vg\kappa$.

 It can be shown that $\v{F}_{e,n}(\vg\kappa, \v{r})$ can be
 orthonormalized such that
 \begin{align}\label{eq:psc_38}
    \int_V \v{F}^\dag_{e,n} (\vg\kappa, \v{r})\cdot \v{F}_{e,n^\prime}
    (\vg\kappa^\prime, \v{r}) d\v{r} =
    \delta_{\vg\kappa,\vg\kappa^\prime} \delta_{nn^\prime}
 \end{align}
 where $n$ indicates the band of the solution for the same
 $\vg\kappa$ as shall be explained.

 In the homogeneous medium case, the transverse modes are degenerate,
 but in a periodic structure considered here, they need not be
 \cite{Busch&John}. Furthermore, the higher frequency modes are
 generated in the Brillouin zone due to the aliasing or wrapping of
 high $\vg\kappa$ modes into the primary zone.  For unlike the
 homogeneous medium case, where $n\in\{1,2\}$, for a fixed
 $\vg\kappa$, there could be many $\kappa_n^2$ for each $\vg\kappa$
 value where $n\in \{1,2,3,\ldots\}$.

 One can also make $\vg\kappa$ countably infinite
 by having the periodic structure nested within a larger periodic
 structure, so that $\vg\kappa$ is discretized by a larger \PBC. One
 can also have only one unit cell within the larger \PBC.  In this
 case, the problem becomes a single region scattering problem within
 a larger \PBC.

 A similar expression to \eqref{eq:dgf_10} for the periodic structure
 case can be derived for the electric dyadic Green's function such
 that
 \begin{align}\label{eq:psc_39}
    \mat{G}_e (\v{r}, \v{r}^\prime) = \sum_{\vg\kappa, n}
    \frac{\v{F}_{e,n}(\vg\kappa, \v{r}) \v{F}_{e,n}^\dag (\vg\kappa,
        \v{r}^\prime)}{\kappa_n^2 - k ^2}
 \end{align}
 Due to reciprocity of the dyadic Green's function, namely that
$\dyad G_e^t(\v r,\v r')=\dyad G_e(\v r', \v r)$,
  it implies that
 $\v F_{e,n}(-\vg\kappa,\v r)=\v F_{e,n}^*(\vg\kappa,\v r)$.

 The magnetic \VWF s can be derived by taking the curl of the
 electric \VWF s. To make them orthonormal \cite{Chew}, one defines
 \begin{align}\label{eq53}
    \v F_{m,n}(\vg\kappa,\v r)=\frac{1}{\kappa_n(\vg\kappa)} \Curl \v
    F_{e,n}(\vg\kappa,\v r)
 \end{align}
 where $m$ above does not indicate an integer but `magnetic'.  It
 satisfies \eqref{eq:psc_36} but with a different boundary condition
 that $\hat n\times \Curl \v F_{m,n}=0$ since $\Curl \v
 F_{m,n}={\kappa_n}\v F_{e,n}$ \footnote{An arbitrary phase factor
    $e^{i\theta}$ can be added to the definition \eqref{eq53}, and the
    orthnormality condition similar to \eqref{eq:psc_38} is still
    preserved.}. The magnetic dyadic Green's function is derived to be
 \begin{align}\label{eq:psc_39a}
    \mat{G}_m (\v{r}, \v{r}^\prime) = \sum_{\vg\kappa, n}
    \frac{\v{F}_{m,n}(\vg\kappa, \v{r}) \v{F}_{m,n}^\dag (\vg\kappa,
        \v{r}^\prime)}{\kappa_n^2 - k ^2}
 \end{align}
 There is a need to include both types of Green's function, and the
 reason will be obvious when one derives the LDOS.
 We define the electric spectral function to be
 \begin{align}\label{eq:ldos_18a}
    \dyad A_e(\v r,\v r')=i\left[\dyad G_e(\v r, \v r')-\dyad
    G_e^\dagger (\v r', \v r) \right]=-2\Im m \left[\dyad G_e(\v r, \v
    r')\right]
 \end{align}

 In a similar manner as before, one defines the spectral function,
 and show that
 \begin{align}\label{eq:tpsc_26}
    \Im m\left(\frac{1}{\kappa_n^2-k^{2}}\right)
    = \frac{\pi E}{2k^2}\left[\delta(E-\mathcal{E} _{\vg\kappa,n}) -
    \delta(E + \mathcal{E}_{\vg\kappa,n})\right]
 \end{align}
 where $\calE=\hbar \kappa_n c$, and $\kappa_n(\vg\kappa)$ which is a
 function of $\vg\kappa$. Therefore,
 \begin{align}\label{eq:tpsc_27}
    \dyad A_e(\v r,\v r',E)= -\frac{\pi E}{k^2}\sum_{{\vg\kappa},n} {\bf
        F}_{e,n}({\vg\kappa},{\bf r}){\bf F}^\dag_{e,n}({\vg\kappa},{\bf
        r}') \delta(E-\mathcal{E}_{{\vg\kappa,n}})
 \end{align}
 A similar expression can be obtained for the magnetic spectral
 function.  From the above, one derives the LDOS to be
 \begin{equation}
 \begin{aligned}
    D(\v r,E)&=-\frac{k^2}{2\pi E} \Tr\left[\dyad A_e(\v r,\v r)+\dyad
    A_m(\v r, \v r)\right]\\
    & = \frac1{2}\sum_{{\vg\kappa},n} \left[
    \left|{\bf F}_{e,n}({\vg\kappa},{\bf r})\right|^2 + \left| {\bf
        F}_{m,n}({\vg\kappa},{\bf r})\right|^2\right]
    \delta(E-\mathcal{E}_{{\vg\kappa,n}})
 \end{aligned}
 \end{equation}
 The above integrates over $\v r$ to become DOS.  One can also
 express the LDOS directly in terms of the Green's dyadics, namely,
 \begin{align}
    D(\v r,E)&=\frac{k^2}{\pi E} \Tr\left\{\Im m\left[ \dyad G_e(\v r,\v
    r)+\dyad G_m(\v r, \v r)\right]\right\}\notag\\
    \tilde D(\v r,\omega)&=\hbar D(\v
    r,\omega)\notag\\&=\frac{\omega\mu_0\epsilon_0}{\pi} \Tr\left\{\Im m\left[
    \dyad G_e(\v r,\v r)+\dyad G_m(\v r, \v r)\right]\right\}
 \end{align}

 The LDOS involves both the electric and magnetic \VWF s.  This is
 because LDOS gives the probability of detecting a photon in a
 complex structure and this probability is proportional to the
 strength of the electromagnetic field.  In a periodic structure
 where standing wave can occur, the strength of the electric field
 may not have the same distribution as the strength of the magnetic
 field.  The above can be integrated over $\v r$ to obtain the DOS.

 At this point, it is expedient to define field correlation functions
 as
 \begin{equation}
 \begin{aligned}\label{eqn60}
    \dyadcal C_e(\v r, \v r',\omega)&=\langle \vcal E(\v r, \omega)\vcal
    E^\dag (\v r',\omega)\rangle\\
    &=-\frac{\omega\mu_0}{\pi } \dyad A_e(\v
    r, \v r',\omega)\Theta(\hbar \omega)\\
    &=\frac{2\omega\mu_0}{\pi }\Im
    m\left[ \dyad G_e(\v r, \v r',\omega)\right]\Theta(\hbar \omega)
 \end{aligned}
 \end{equation}
 \begin{equation}
 \begin{aligned}\label{eqn61}
    \dyadcal C_m(\v r, \v r',\omega)&=\langle \vcal H(\v r,\omega)\vcal
    H^\dag (\v r',\omega)\rangle\\
    &=-\frac{\omega\epsilon_0}{\pi } \dyad
    A_m(\v r, \v r',\omega)\Theta(\hbar
    \omega)\\
    &=\frac{2\omega\epsilon_0}{\pi } \Im m\left[ \dyad G_m(\v r,
    \v r',\omega)\right]\Theta(\hbar \omega)
 \end{aligned}
 \end{equation}
 Then the LDOS can be related to these correlation functions as
 \begin{align}\label{eqn62}
    \tilde D(\v r, &\omega)\Theta(\hbar
    \omega)d\omega\notag\\&=\Tr\left[\frac{1}{2} \epsilon_0 \dyadcal C_e(\v r,\v
    r,\omega)+\frac{1}{2} \mu_0 \dyadcal C_m(\v r,\v
    r,\omega)\right]d\omega
 \end{align}
 In a homogeneous medium case, $\epsilon_0 \dyadcal C_e(\v r,\v
 r)=\mu_0\dyadcal C_m(\v r,\v r)$, and the above reduces to
 \eqref{eqn40}.

 In general, $\kappa_n(\vg\kappa)=\kappa_n(k,\theta_k,\phi_k)$ is
 dependent on the direction of $\vg\kappa$, or anisotropic. This is
 unlike the homogeneous medium case, $\kappa_n(\vg\kappa)=k$, where
 it is much simpler or isotropic. Therefore, when loss is included,
 the prove of Equation \eqref{eq:lbal_40}, valid for homogeneous
 medium, is more difficult for normalization now. But a general
 spectral function with line-broadening, local density of states, and
 density of states equations can be assumed, as shall be confirmed later.  Moreover, \eqref{eqn60} and
 \eqref{eqn61} can be confirmed in Section \ref{sect8} and are
 commensurate with FDT.

 \section{Inhomogeneous Medium Case}

 \subsection{Lossless, Reciprocal Case}

 In this case, the relative permittivity and permeability tensors
 $\dyadg\epsilon_r(\v r)=\dyadg\epsilon(\v r)/\epsilon_0$ and
 $\dyadg\mu_r(\v r)=\dyadg\mu(\v r)/\mu_0$, respectively. They are
 Hermitian tensors \cite{Kong}. The electric dyadic Green's function
 satisfies the equation
 \begin{align} \label{eq:limc_1}
    \nabla\times\dyadg\mu^{\,-1}_r(\v r)\cdot \nabla\times\overline{\bf G}_e({\bf r},{\bf r}')&-k^2\dyadg\epsilon_r(\v r)\cdot\overline{\bf G}_e({\bf r},{\bf r}')\notag\\&
    =\overline{\bf I}\delta({\bf r}-{\bf r}')
 \end{align}
 An electric \VWF\ is defined to be the solution to
 \begin{align}\label{eq:limc_2}
    \nabla\times\dyadg\mu^{\,-1}_r(\v r)\cdot \nabla\times{\bf F}_{e,n}({\vg\kappa},{\bf r})-\kappa_n^2\dyadg\epsilon_r(\v r)\cdot{\bf F}_{e,n}({\vg\kappa},{\bf r})=0
 \end{align}
 %It is to be noted that $\dyadg\mu_r(\v r)$ and $\dyadg\epsilon_r(\v
 %r)$ may be functions of frequency $\omega$.  But for a fixed
 %$\omega$, the above system is Hermitian and self-adjoint.  Hence,
 %the eigenvalues $\kappa_n^2$ are real, and the eigenfunctions are
 %complete.

 The medium can be arranged as a periodic structure, and the
 derivation of the eigenfunction above is guided by the Bloch-Floquet
 theorem.  The eigenvalue $\kappa_n(\vg\kappa)$ is then a function of
 $\vg\kappa$. To this end, one defines the eigenfunctions $\v
 F_{e,n}(\vg\kappa,\v r)$ similar to the periodic structure case as
 in \eqref{eq:psc_37}. But the Bloch-Floquet modes are $\dyadg
 \epsilon_r(\v r)$ orthonormal since we can prove from
 \eqref{eq:limc_2} that
 \begin{align}\label{eqn65}
    \int_V d\v r \v F_{en'}^\dag(\vg\kappa,\v r) \cdot\dyadg
    \epsilon_r(\v r) \cdot \v F_{e,n}(\vg\kappa',\v
    r)=\delta_{nn'}\delta_{\vg\kappa, \vg\kappa'}
 \end{align}
 Using the above, it can be shown that
 \begin{align}
    \dyad I \delta(\v r-\v r')&=\sum_{\vg\kappa, n} \v
    F_{e,n}(\vg\kappa,\v r) \v
    F_{e,n}^\dagger (\vg\kappa,\v r')\cdot\dyadg\epsilon_r(\v r')\nonumber\\
    &=\sum_{\vg\kappa, n} \dyadg\epsilon_r(\v r)\cdot \v
    F_{e,n}(\vg\kappa,\v r) \v F_{e,n}^\dagger (\vg\kappa,\v r')\nonumber\\
    &
    =\sum_{\vg\kappa, n} \dyadg\epsilon_r^{\frac{1}2}(\v r)\cdot \v
    F_{e,n}(\vg\kappa,\v r) \left[\dyadg\epsilon_r^{\frac{1}2}(\v
    r')\cdot\v F_{e,n}(\vg\kappa,\v r')\right]^\dag
 \end{align}
 %The above identities can be easily proven from \eqref{eqn65}.
 %
 The electric dyadic Green's function, despite the change in the
 orthonormality condition expressed by \eqref{eqn65}, can be defined
 as before in \eqref{eq:psc_39}, and the electric spectral function
 can be similarly defined.

 Similarly, a magnetic dyadic Green's function can be defined as
 \begin{align} \label{eq:limc_1a}
    \nabla\times\dyadg\epsilon^{\,-1}_r(\v r)\cdot \nabla\times\overline{\bf G}_m({\bf r},{\bf r}')&
    -k^2\dyadg\mu_r(\v r)\cdot\overline{\bf G}_m({\bf r},{\bf r}')\notag\\&=\overline{\bf I}\delta({\bf r}-{\bf r}')
 \end{align}
 with its magnetic \VWF\ to be the solution to
 \begin{align}\label{eq:limc_2a}
    \nabla\times\dyadg\epsilon^{\,-1}_r(\v r)\cdot \nabla\times{\bf F}_{m,n}({\vg\kappa},{\bf r})-\kappa_n^2\dyadg\mu_r(\v r)\cdot{\bf F}_{m,n}({\vg\kappa},{\bf r})=0
 \end{align}
 and its corresponding magnetic spectral function.

 The LDOS is related to the spectral functions as
 %\begin{widetext}
 \begin{align}
    \tilde D(\v r,\omega)=\hbar D(\v r,
    E)&=-\frac{\omega\mu_0\epsilon_0}{2\pi }
    \Tr\left[\dyadg\epsilon_r^{\frac12}(\v r)\cdot\dyad A_e(\v r,\v
    r)\cdot\dyadg\epsilon_r^{\frac12}(\v r)\right.\notag\\&\left.+ \dyadg\mu_r^{\frac12}(\v
    r)\cdot\dyad A_m(\v r,\v r)\cdot\dyadg\mu_r^{\frac12}(\v r) \right]
 \end{align}
 %\end{widetext}
 The modification is necessary because of the new orthonormality
 condition expressed by \eqref{eqn65}.
 Upon substituting \eqref{eq:psc_39} and \eqref{eq:psc_39a} into the
 above, invoking the property of the vector wave function that
 follows \eqref{eq:psc_39},
  and going through the manipulation as in Section \ref{sect4},
 equations \eqref{eq:ldos_21} to \eqref{eq:ldos_27}, finally, the
 LDOS can be shown to be modified as
%\begin{widetext}
 \begin{align}
    D(&\v r,E)=\frac{1}2 \sum_{\vg\kappa,n} \left[\v
    F_{e,n}^\dag(\vg\kappa,\v r) \cdot\dyadg \epsilon_r(\v r) \cdot \v
    F_{e,n}(\vg\kappa,\v r)\right.\notag\\&\left.+ \v F_{m,n}^\dag(\vg\kappa,\v r) \cdot\dyadg
    \mu_r(\v r) \cdot \v F_{m,n}(\vg\kappa,\v r) \right] \delta(E-\calE)
 \end{align}
%\end{widetext}
 By using the orthonormality condition \eqref{eqn65}, the above
 integrates over $\v r$ to become the DOS.   The LDOS can also be
 written directly in terms of the dyadic Green's functions as
\begin{widetext}
 \begin{align} \label{eq:66}
    \tilde D(\v r, \omega)=\hbar D(\v r,E)=\frac{
        \omega\mu_0\epsilon_0}{\pi } \Tr\left\{\dyadg\epsilon_r^{\frac12}(\v
    r)\cdot \Im m\left[ \dyad G_e(\v r,\v
    r)\right]\cdot\dyadg\epsilon_r^{\frac12}(\v r)+
    \dyadg\mu_r^{\frac12}(\v r)\cdot\Im m\left[\dyad G_m(\v r,\v
    r)\right]\cdot\dyadg\mu_r^{\frac12}(\v r) \right\}
 \end{align}
\end{widetext}

 With this modification, the photon energy density can be calculated
 accordingly.  Notice that the LDOS becomes large when
 $\dyadg\epsilon_r(\v r)$ or $\dyadg\mu_r(\v r)$ is large.
 %It is to
% be noted that the above analysis remains valid even if
% $\dyadg\epsilon_r(\v r)$ or $\dyadg\mu_r(\v r)$ are functions of
% $\omega$ as long as we fix $\omega$ during the analysis.
% However,
Since
 the Kramers-Kronig relation \cite{Kramers&Kronig} implies that a
 frequency dependent or dispersive medium is also necessary lossy,
 we assume that the medium is frequency independent.

 In the above, we can similarly define correlation functions for the
 electric and magnetic fields as in \eqref{eqn60} and \eqref{eqn61}.
 Then we can relate them to the LDOS as
%\begin{widetext}
 \begin{align}\label{eqn62a}
    \tilde D(\v r, &\omega)\Theta(\hbar
    \omega)d\omega=\Tr\left[\frac{1}{2} \epsilon_0
    \dyadg\epsilon_r^\frac{1}{2}(\v r)\cdot\dyadcal C_e(\v r,\v
    r,\omega)\cdot\dyadg\epsilon_r^\frac{1}{2}(\v
    r)\right.\notag\\&\left.
    +\frac{1}{2}\mu_0
    \dyadg\mu_r^\frac{1}{2}(\v r)\cdot\dyadcal C_m(\v r,\v
    r,\omega)\cdot\dyadg\mu_r^\frac{1}{2}(\v r)\right]d\omega
 \end{align}
%\end{widetext}

 \subsection{Lossy, Reciprocal Case}

 The equation for the electric dyadic Green's function in this case
 is the same as before. The compendium equation for the vector wave
 function can be defined as before in \eqref{eq:limc_2} and
 \eqref{eq:limc_2a}, but now, $\dyadg \mu_r(\v r)$ and $\dyadg
 \epsilon_r(\v r)$ are non-Hermitian, but symmetric tensors to
 represent a general lossy, reciprocal medium. In this case, the
 auxiliary or transpose equations have to be defined and
 $\kappa_n(\vg\kappa)$ is not a real number anymore
 \cite{Chew,DaiEtal}.
 %\footnote{For the case of non-reciprocal medium,
 %the auxiliary problem may involves a different medium, making the
 %derivation more involved.}
 One can assume that the lossy medium
 forms periodic islands immersed in the lossless background. Then one
 can still use the Bloch-Floquet theorem as guidance in deriving the
 eigensolution. Again, one can also use \PBC\ to make $\vg\kappa$
 countably infinite.

 The Bloch-Floquet modes are $\dyadg \epsilon_r(\v r)$ orthonormal
 since it can be shown that
 \begin{align}\label{eq:lrimc_1}
    \int_V d\v r \v F_{e,n,a}^\dag(\vg\kappa,\v r) \cdot\dyadg \epsilon_r(\v
    r) \cdot \v F_{en'}(\vg\kappa',\v r)=\delta_{nn'}\delta_{\vg\kappa,
        \vg\kappa'}
 \end{align}
 where $\v F_{e,n,a}$ is the vector wave function of an auxiliary
 problem where material properties $\epsilon_r(\v r)$ and $\mu_r(\v r)$ are replaced by
 their conjugate transpose, implying an active medium.  It can be
 shown that $\v F_{e,n,a}=\v  F^*_{e,n}$ in this case.
%
%
% Notice that the $\dag$ is now replaced by $t$ for the orthonormality
% condition because this is the lossy reciprocal case.
 %
 It can be shown that the electric dyadic Green's function can be
 expanded as
 \begin{align}\label{eq:lrimc_39}
    \mat{G}_e (\v{r}, \v{r}^\prime) = \sum_{\vg\kappa, n}
    \frac{\v{F}_{e,n}(\vg\kappa, \v{r}) \v{F}_{e,n,a}^{\dag }(\vg\kappa,
        \v{r}^\prime)}{\kappa_n^2 - k ^2}
 \end{align}
 The above assumes the completeness of the eigenmodes even when the
 medium is lossy. The above is also analogous to the complex
 symmetric generalized eigenvalue matrix systems in linear algebra
 where the right and left eigenvectors are orthogonal to each other
 via a matrix.\footnote{If the auxiliary equation is chosen such that
    its medium is described by the $\dyadg\epsilon^\dag$ and
    $\dyadg\mu^\dag$, namely, the conjugate transpose of the original
    problem corresponding to an active medium, then the $\v F_{e,n}^t$
    above can be replaced by $\v F_{a,e,n}^\dag$ where $\v F_{a,e,n}$ is
    the solution of the auxiliary equation.}

 As before, the electric spectral function is defined to be
 \begin{align}\label{eq:lrimc_18}
    \dyad A_e(\v r,\v r')=i\left[\dyad G_e(\v r, \v r')-\dyad
    G_e^\dagger (\v r', \v r)\right]=-2\I  \left[\dyad G_e(\v r, \v
    r')\right]
 \end{align}
 The above form is more complicated compared to the lossless
 inhomogeneous medium case.
 Here, $\kappa_n^2$ is complex in \eqref{eq:lrimc_39}, and we see
 that the above gives rise to line broadening as before.  Moreover,
 at one given frequency $\omega$, a cluster of modes is excited.
 However, as shall be shown later, the spectral function can be
 related to the correlation function of the field, and hence, LDOS.

 The permittivity and permeability tensors are of the form
 \begin{align}\label{eq:lrimc_4}
    \dyadg\epsilon=\dyadg\epsilon'+i\dyadg\epsilon'',\qquad
    \dyadg\mu=\dyadg\mu'+i\dyadg\mu''
 \end{align}
 where both the real and imaginary parts of the above tensors are
 real symmetric for reciprocal media and hence are Hermitian. For
 dispersive media,
 one can
 define effective permittivity and permeability tensors to be
 \cite{Haus,Loudon,Ruppin}
 \begin{align}\label{eq:lrimc_5}
    \dyadg\epsilon_e=\frac {d\omega\dyadg\epsilon'}{d\omega},\qquad
    \dyadg\mu_e=\frac {d\omega\dyadg\mu'}{d\omega}
 \end{align}
 where the above tensors are Hermitian.  Since the above is derived
 using perturbation argument, it is only valid for the low loss case.
 Similar to the lossless inhomogeneous case, the generalized LDOS is
 defined as
 %\begin{widetext}
 \begin{align}\label{eq:lrimc_6}
    \tilde D(\v r, \omega)&=\hbar D(\v r,E)\notag\\&
    =\frac{ \omega}{\pi }
    \Tr\left\{\mu_0\dyadg\epsilon_e^{\frac12}(\v r)\cdot \Im m\left[
    \dyad G_e(\v r,\v r)\right]\cdot\dyadg\epsilon_e^{\frac12}(\v r)\right.\notag\\&\quad\quad\left.+
    \epsilon_0\dyadg\mu_e^{\frac12}(\v r)\cdot\Im m\left[\dyad G_m(\v
    r,\v r)\right]\cdot\dyadg\mu_e^{\frac12}(\v r) \right\}
 \end{align}
%\end{widetext}
 The above LDOS is not provable for a general inhomogeneous lossy
 medium, because the line broadening is rather complicated here. But
 we will justify it in the next section.
 %
 %Hence, the assumption above is that the integration ($\sum_{k,n}$)
 %over the cluster of modes associated with the peak $k=k$ still
 %represents the DOS or LDOS at that frequency.

 The Kramers-Kronig relations \cite{Kramers&Kronig} requires that a
 lossy medium is also dispersive, and the lossless medium be
 non-dispersive.  Hence, the above reduces to \eqref{eq:66} in the
 lossless medium case.
 The above analysis remains valid even if $\dyadg\epsilon_r(\v r)$
 or $\dyadg\mu_r(\v r)$ are functions of $\omega$.  As a consequence,
 the eigenvalue $\kappa_n(\vg\kappa,\omega)$ is a function of both
 $\vg\kappa$ and $\omega$, so are the eigenfunctions $\v
 F_n(\vg\kappa,\omega,\v r)$ of the electric and magnetic types.
 Equations \eqref{eq:lrimc_1} and \eqref{eq:lrimc_39} are valid for
 dispersive media as long as $\omega$ is fixed.

 \section{Spectral Function and Field Correlation}\label{sect8}

 For the lossless media, we have shown that the spectral functions is
 related to the local density of states, and hence, they can be
 related to the field correlation functions.  This is because the
 self correlation of the field is related to local energy density,
 and hence, the local density of states.

 For the lossy media case, it is not clear that the spectral function
 is related to the local density of states, but we can relate the
 spectral function to field correlation functions directly, as we
 shall show in this section.  In this manner, the spectral function
 can be related to local energy density and hence, the local density
 of states.

 %\section{Fluctuation Dissipation Theorem}
 %
 %
 %
 %Fluctuation dissipation theorem describes the energy balance of a
 %lossy system with random thermal sources of its environment, known
 %as Langevin sources. As much energy is radiated into the environment
 %as it is absorbed from the environment.  Hence, the simple harmonic
 %oscillators in the environment are not damped and therefore, they
 %can be connected to quantum harmonic oscillators. To see the meaning
 %of the spectral function derived previously for lossy media in the
 %previous section, it will be connected to the field correlation
 %function via the fluctuation dissipation theorem.

 \subsection{Statistical Average of Random Fields}

 Before proceeding with this section, one needs to define some
 properties of the correlations of random fields and sources. Imagine
 that random Langevin sources are generating random fields due to
 thermal excitation of these sources by the environment.
 These random fields are stationary time processes, namely the time
 correlations of these fields depend only on time delay between the
 fields, or $t-t'$ \cite{Landau&Lifshitz,JoulainEtal}. Then the
 frequency correlations between these fields depend only on
 $\delta(\omega-\omega')$, namely, they are uncorrelated in
 frequency.
 When the Langevin sources are in thermal equilibrium with its
 environment, and as shall be shown, the correlation of the fields at
 different locations $\v r$ and $\v r'$ are related via the spectral
 function.

 If the Fourier transform of the random field $\v E(\v r, t)$ is
 defined as
 \begin{align}
    \v E(\v r, \omega)=\int_{-\infty}^{\infty} dt \v E(\v r,
    t)e^{i\omega t}
 \end{align}
 then the $\v E(\v r,\omega)$ has the dimension $1/\omega$ times
 field.  The general form of the correlation of the electric fields
 at two locations $\v r$ and $\v r'$ is
 \begin{align}\label{FDT:eq77a}
    \langle \v E(\v r,\omega)\v E^\dag(\v
    r',\omega')\rangle&=\delta(\omega-\omega') \langle \vcal E(\v
    r,\omega)\vcal E^\dag(\v
    r',\omega')\rangle \nonumber\\&=\delta(\omega-\omega')\dyadcal C_e(\v r,\v
    r',\omega)
 \end{align}
 Similarly, the correlation function for the magnetic field is of the
 form
 \begin{align}\label{FDT:eq78a}
    \langle \v H(\v r,\omega)\v H^\dag(\v
    r',\omega')\rangle&=\delta(\omega-\omega')\langle \vcal H(\v
    r,\omega)\vcal H^\dag(\v
    r',\omega')\rangle\nonumber\\
    &=\delta(\omega-\omega')\dyadcal C_m(\v r,\v
    r',\omega)
 \end{align}
 In the above $\dyadcal C_e$ and $\dyadcal C_m$ are proportional to
 energy density per unit radian frequency.

 \subsection{Connection of the Spectral Function to Medium Loss}

 The random fields in a medium at thermal equilibrium are generated
 by Langevin current sources \cite{DungEtal}. To this end, one
 rewrites electric dyadic Green's function equation in operator
 notation as:
 \begin{equation}
    \left(\overline{\boldsymbol{\mathcal{D}}} - {k}^2
    \dyadcal{E}_r\right) \dyadcal{G}_e = \dyadcal{I}
 \end{equation}
 where
 \begin{align} \dyadcal D \Rightarrow
    \left[\nabla\times\dyadg\mu_r^{\,-1}(\v r)\nabla\times\right],
    \qquad \dyadcal{E}_r \Rightarrow \left[\dyadg\epsilon_r(\v r)\right]
 \end{align}
 These are the operators (or general Hilbert space representation of
 the operators) originally defined in coordinate space.  For lossy
 media, $\dyadcal D$ and $\dyadcal {E}_r$ are non Hermitian, and the
 expressions right of the arrows are their coordinate space
 representations.

 Equivalently, the Green's dyadic operator is then
 \begin{equation}
    \dyadcal{G}_e = \left(\dyadcal{D} - {k}^2 \dyadcal{E}_r \right)^{-1}
 \end{equation}
 We assume that the relevant boundary condition or radiation
 condition is imposed so that the inverse of the operator above is
 uniquely defined.
 Using these notations, an interesting expression for the spectral
 function operator can be derived.  It can be shown that
 \begin{align}\label{eqn85}
    \dyadg{\Gamma}&=i\left[\left({\dyadcal {G}_e^{a}}\right)^{-1} -
    \dyadcal{G}_e^{-1}\right] = i\left[\dyadcal{D}^a-\dyadcal{D}-{k}^2
    \left(\dyadcal{E}_r^{a} - \dyadcal{E}_r\right)\right]\nonumber\\& =2\Im
    m\left(\dyadcal D\right) - 2  {k}^2 \Im m\left(\dyadcal{E}_r\right)
 \end{align}
 The above $\dyadg{\Gamma}$ would be zero for a lossless medium for
 which $\dyadg\mu_r(\v r)$ and $\dyadg\epsilon_r(\v r)$ are
 Hermitian.  Hence $\dyadg{\Gamma}$ represents the loss of the
 system.  In the above, the adjoint operator denoted by superscript
 $a$, is as defined in \eqref{eqn22} where the 3-vector and Hilbert
 space nature of the space is accounted for.

 Multiplying the above by $\dyadcal{G}_e$ and $\dyadcal{G}_e^a$ from
 the left and right, respectively, the electric spectral function
 operator is given by:
 \begin{equation}\label{eq:ldos73}
    \dyadcal{A}_e= i \left(\dyadcal{G}_e - \dyadcal{G}_e^a\right) =
    \dyadcal G_e\,\dyadg \Gamma\,\dyadcal G_e^a
 \end{equation}
 The above can also be written more suggestively so that
 \begin{equation}\label{eqn87}
    \left(\dyadcal{D} - {k}^2 \dyadcal{E}_r \right)\dyadcal{A}_e= \dyadg
    \Gamma\,\dyadcal G_e^a
 \end{equation}
 It implies that the spectral function operator $\dyadcal{A}_e$ is
 generated by distributed sources on the \RHS\ that are related to
 the Langevin sources.
 The right-hand side of the above can be interpreted as induced
 sources due to the back propagation of field via $\dyadcal G_e^a$
 into the lossy part of the medium represented by $\dyadg\Gamma$.

 However, the spectral function operator can also be defined more
 simply as $\dyadcal A_e=i(\dyadcal G_e-\dyadcal G_e^a)$.  As shall
 be shown, the spectral function is proportional to the field
 correlation function, and it appears that the fields are correlated
 by the difference of a forward Green's function and a backward
 Green's function, making it appear like a radiating source occurring
 concurrently with an absorbing source.

 It is to be noted that in \eqref{eq:ldos73}, the operators in the
 3-vector functional Hilbert space, and the $\dag$ sign implies the
 adjoint operator defined in this infinite dimensional space. With
 this caution, the coordinate representation of the spectral function
 operator $\dyadcal A_e$ in \eqref{eq:ldos73} becomes
 \begin{widetext}
 \begin{align}\label{eq:ldos74}
    \dyad A_e(\v r, \v r')=\dyadg \epsilon_r^{-1}(\v r)\int d\v r''
    \Curl \dyad G_m(\v r,\v r'')\cdot \dyadg \mu_r(\v
    r'')\cdot\left[2\Im m \dyadg \mu_r^{-1}(\v r'')\right] \cdot \nabla
    ''\times \dyad G_e^\dagger(\v r', \v r'')]\notag\\
    -\int d\v r''
    \dyad G_e(\v r,\v r'')\cdot\left[ 2k^2\Im m \dyadg \epsilon_r(\v
    r'')\right]\cdot \dyad G_e^\dagger(\v r', \v r'')
 \end{align}
 \end{widetext}
 In the above, one has made use of that
 \begin{align}\label{FDT:eq87}
    \left[\dyadg \mu_r^{-1}(\v r)\cdot\Curl \dyad G_e(\v r, \v
    r')\right]^t=\dyadg \epsilon_r^{-1}(\v r')\cdot\nabla '\times \dyad
    G_m(\v r',\v r)
 \end{align}
 The above can be derived using reciprocity
 theorem \footnote{Derivation available upon request.}. It is the
 generalization of (1.4.14b) in \cite{Chew} to inhomogeneous,
 anisotropic media. Furthermore, it can be shown that
 \begin{align}\label{FDT:eq_95}
    -\dyadg \mu_r \Im m \left( \dyadg \mu_r^{-1}\right) \dyadg \mu_r
    ^\dag= \Im m \left(\dyadg\mu_r\right)
 \end{align}
 Therefore, \eqref{eq:ldos74} can be rewritten as
 \begin{widetext}
 \begin{align}\label{eqn91}
    \dyad A_e(\v r, \v r')=-2\Im m \dyad G_e(\v r, \v r')=-\dyadg
    \epsilon_r^{-1}(\v r)\int d\v r'' \Curl \dyad G_m(\v r,\v r'')\cdot
    \left[2\Im m \dyadg \mu_r(\v r'')\right] \cdot \left[\dyadg
    \mu_r^{-1}(\v r'')\right]^\dag\cdot \nabla
    ''\times \dyad G_e^\dagger(\v r', \v r'')\nonumber\\
    -\int d\v r''
    \dyad G_e(\v r,\v r'')\cdot\left[ 2k^2\Im m \dyadg \epsilon_r(\v
    r'')\right]\cdot \dyad G_e^\dagger(\v r', \v r'')
 \end{align}
 \end{widetext}

 \subsection{Connection of the Field Correlations to Langevin Sources}

 Alternatively, \eqref{eqn91} above can be related to the
 correlations of electric fields. To this end, one defines electric
 field due to the electric Langevin sources as
 \begin{align}
    \v E_e(\v r,\omega) = i\omega \mu_0 \int d\v r' \dyad G_e(\v r, \v
    r', \omega)\cdot \v J(\v r',\omega)
 \end{align}
 The average  of the outer product of electric fields due to the
 electric type sources at two different frequencies is then given by
 the field correlation
 %\begin{widetext}
 \begin{align}\label{eq:FDT_89}
    \langle\v E_e(\v r,\omega)&\v E_e^\dag (\v r',\omega')\rangle
    =\omega^2\mu_0^2 \int d \v r ''\int d\v r''' \dyad G_e(\v r, \v
    r'',\omega)\notag\\&\cdot\langle \v J(\v r'',\omega) \v J^\dag (\v
    r''',\omega')\rangle\cdot \dyad G_e^\dag(\v r', \v r''',\omega')
 \end{align}
 %\end{widetext}
 Similarly, one defines the electric field due to magnetic Langevin
 sources as
 \begin{align}\label{eq:FDT_90}
    \v E_m(\v r,\omega) = \dyadg\epsilon_r^{-1} \int d\v r''
    \nabla\times \dyad G_m(\v r, \v r'', \omega)\cdot \v M(\v
    r'',\omega)
 \end{align}
 The average of outer product of electric fields due to the magnetic
 type sources is then
 \begin{widetext}
 \begin{align}\label{FDT:eq_91}
    \langle\v E_m(\v r,\omega)\v E_m^\dag (\v r',\omega')\rangle =\dyadg
    \epsilon_r^{-1}(\v r) \int d \v r_a \int d\v r_b\Curl \dyad G_m(\v
    r, \v r_a,\omega)\cdot\langle \v M(\v r_a,\omega) \v M^\dag (\v
    r_b,\omega')\rangle
    \cdot{\left[\dyadg\mu_r^{-1}(\v r_b)\right]^\dag}\cdot
    \nabla_b\times \dyad G_e^\dag(\v r', \v r_b,\omega)
 \end{align}
\end{widetext}
 In the above, one has made use of \eqref{FDT:eq87}, the fact that
 $\dyad G^t(\v r,\v r')=\dyad G(\v r',\v r)$ from reciprocity,
 and that $\dyadg \mu_r$ is a
 symmetric tensor. Consequently, assuming that the field due to
 electric Langevin sources is uncorrelated to the field due to
 magnetic Langevin sources, we have
 \begin{widetext}
 \begin{align}\label{eqn96}
    &\langle\v E(\v r,\omega)\v E^\dag (\v r',\omega')\rangle=\langle\v
    E_e(\v r,\omega)\v E_e^\dag (\v r',\omega')\rangle+\langle\v E_m(\v
    r,\omega)\v E_m^\dag (\v r',\omega')\rangle
    =\omega^2\mu_0^2 \int d \v r _a\int d\v r_b \dyad G_e(\v r, \v
    r_a,\omega)\cdot\langle \v J(\v r_a,\omega) \v J^\dag (\v
    r_b,\omega')\rangle\notag\\&\cdot \dyad G_e^\dag(\v r', \v
    r_b,\omega')
    +\dyadg \epsilon_r^{-1}(\v r)\cdot \int d \v r_a \int
    d\v r_b\Curl \dyad G_m(\v r, \v r_a,\omega)\cdot\langle \v M(\v
    r_a,\omega) \v M^\dag (\v r_b,\omega')\rangle
    \cdot{\left[\dyadg\mu_r^{-1}(\v r_b)\right]^\dag}\cdot
    \nabla_b\times \dyad G_e^\dag(\v r', \v r_b,\omega)
 \end{align}
 \end{widetext}
 Notice that the above is of the same structural form as
 \eqref{eqn91}.

 \subsection{Connection of Spectral Function to the Field
    Correlations}

 %
 %\begin{align}\label{FDT:eq_93}
 % \langle \v J(\v r, \omega)\v J^\dag (\v r',\omega')\rangle
 % =\delta(\omega-\omega')\delta(\v r-\v r')
 %\dyadcal J(\v r, \omega)
 % \end{align}
 % %
 %\begin{align}\label{FDT:eq_96}
 % \langle \v M(\v r, \omega)\v M^\dag (\v r',\omega')\rangle
 % =\delta(\omega-\omega')\delta(\v r-\v r')
 %\dyadcal M(\v r, \omega)
 %\end{align}
 %The above correlations of the Langevin source with $\delta(\v r-\v
 %r')$ imply that only the fields from the same Langevin source will
 %be correlated.
 %

 By comparing \eqref{eqn91} and \eqref{eqn96}, and in order for them
 to be proportional to each other, it implies that
 the Langevin currents have
 correlations of the form
 \begin{align}\label{FDT:eq_93a}
    \langle \v J(\v r, \omega)\v J^\dag (&\v r',\omega')\rangle
    \notag\\&=\delta(\omega-\omega')\delta(\v r-\v r')
    \frac{\omega\epsilon_0}{\pi} \Im m\left[\dyadg \epsilon_r (\v
    r)\right]\Theta(\hbar \omega)
 \end{align}
 and
 \begin{align}\label{FDT:eq_94a}
    \langle \v M(\v r, \omega)\v M^\dag (&\v r',\omega')\rangle
    \notag\\&=\delta(\omega-\omega')\delta(\v r-\v r')
    \frac{\omega\mu_0}{\pi} \Im m\left[\dyadg \mu_r (\v
    r)\right]\Theta(\hbar \omega)
 \end{align}
 The above implies that Langevin sources are uncorrelated in frequency and
 space.  It also implies that the field correlation function in \eqref{eqn96} becomes
 \begin{widetext}
 \begin{align}\label{FDT:eq_91a}
    \langle\v E(\v r,\omega)\v E^\dag (\v r',\omega')\rangle
    =\delta(\omega-\omega')\left\{\dyadg \epsilon_r^{-1}(\v r) \int d \v
    r'' \Curl \dyad G_m(\v r, \v r'',\omega)\cdot
    \frac{\omega\mu_0}{\pi} \Im m\left[\dyadg \mu_r (\v
    r)\right]\Theta(\hbar \omega)
    %\dyadcal M(\v r'',
    %\omega)
    \cdot{\left[\dyadg\mu_r^{-1}(\v r'')\right]^\dag}\cdot \right. \notag\\
    \nabla''\times \dyad G_e^\dag(\v r', \v r'',\omega)
    + \left. \omega^2\mu_0^2 \int d \v r '' \dyad G_e(\v r, \v
    r'',\omega) \cdot \frac{\omega\epsilon_0}{\pi} \Im m\left[\dyadg
    \epsilon_r (\v
    r)\right]\Theta(\hbar \omega)
    %\dyadcal J(\v r'',\omega)
    \cdot \dyad G_e^\dag(\v r', \v r'',\omega') \right\}
 \end{align}
 \end{widetext}
 The above implies that the right-hand side is proportional to the
 spectral function times $\Theta(\hbar \omega)$.  Consequently, the
 field correlation and the spectral function are related as indicate
 before in equations \eqref{eqn39}, \eqref{eqn60}, and \eqref{eqn61}.
 Namely,
 \begin{widetext}
 \begin{align}\label{eqn97}
    \langle \v E(\v r,\omega)\v E^\dag(\v
    r',\omega')\rangle=\delta(\omega-\omega')\langle \vcal E(\v
    r,\omega)\vcal E^\dag(\v r',\omega')\rangle
    =\delta(\omega-\omega')\dyadcal C_e(\v r,\v
    r',\omega)=\delta(\omega-\omega')\frac{ \omega\mu_0}{\pi }
    \Im m\left[ \dyad G_e(\v r,\v
    r')\right] \Theta(\hbar \omega)
 \end{align}
 \end{widetext}
 Or that $\dyadcal C_e(\v r,\v r',\omega)$ is connected to the
 electric spectral functions as
 \begin{align}\label{FDT:eq78}
    \dyadcal C_e(\v r,\v r',\omega)=\frac{ \omega\mu_0}{\pi }
    \Im m\left[ \dyad G_e(\v r,\v
    r')\right] \Theta(\hbar \omega)
 \end{align}

 The assertions in equations \eqref{FDT:eq_93a} and
 \eqref{FDT:eq_94a} are justified for the following reasons:

 \begin{itemize}

    \item The spectral function consists of two bilinear terms, one of
    which is linearly proportional to the permittivity loss and the
    other one linearly proportional to the permeability loss.  We can
    also derive \eqref{FDT:eq_93a} and \eqref{FDT:eq_94a} by turning off
    the loss of each kind respectively and equating \eqref{eqn91} and
    \eqref{eqn96} with the appropriate proportionality constant.

    \item  The Green's dyadic is undefined for a lossless medium unless
    infinitesimal small loss is introduced.  Hence, one thinks of
    \eqref{eqn96}, and hence, \eqref{FDT:eq_91a} being valid for the
    case of infinitesimally small loss media as well. Therefore,
    \eqref{FDT:eq78} is valid for the lossless case as well.

    \item For the lossless case, \eqref{eq:ped_31}, \eqref{eqn60} and
    \eqref{eqn61} are confirmations that the above derivation leading to
    \eqref{FDT:eq78} is correct.  Furthermore, the spectral function can
    be related to the local density of states, and hence energy density
    after multiplying by $\Theta(\hbar \omega)$. This is further
    affirmation that Equation \eqref{FDT:eq78} is at least valid when
    $\v r=\v r'$.

 \end{itemize}

  A correlation function for the magnetic field can similarly be defined
  as
  \begin{align}
    \langle \v H(\v r,\omega)\v H^\dag(\v
    r',\omega')\rangle=\delta(\omega-\omega')\dyadcal C_m(\v r,\v
    r',\omega)
  \end{align}
  where $\dyadcal C_m(\v r,\v r',\omega)$ can be connected to the
  magnetic spectral functions as
  \begin{align}
    \dyadcal C_m(\v r,\v r',\omega)=\frac{ \omega\epsilon_0}{\pi }
    %\frac{1}{\epsilon}\dyadg\epsilon_e^{\frac12}(\v r)\cdot \Im
    %m\left[ \dyad G_e(\v r,\v
    %r')\right]\cdot\dyadg\epsilon_e^{\frac12}(\v r')
    %+
    \Im m\left[\dyad G_m(\v r,\v r')\right]\Theta(\hbar \omega)
  \end{align}
  The above results are also corroborated by the fluctuation
  dissipation theorem \cite{Landau&Lifshitz,JoulainEtal}.

  \subsection{Physical Interpretation}

  Now, however, \eqref{eqn91} and \eqref{FDT:eq_91a} have a nice
  physical interpretation. It implies that there are two types of
  contributions to the electric field correlation function or the
  electric spectral function: one that comes from magnetic type
  Langevin current sources and the second that comes from electric
  type Langevin current sources.  Since the complex conjugation of a
  field corresponds to time reversal or back propagation, a field at
  point $\v r'$ back propagates into the lossy medium to the source
  point at $\v r''$ via the backward Green's function $\dyad
  G_e^\dag$. The Langevin source at point $\v r''$ can only correlate
  or be coherent with the same Langevin source. The same Langevin
  source radiates a field that propagates from the source point $\v
  r''$ to the field point at $\v r$ via the forward Green's function
  $\dyad G_e$. Different Green's functions are used depending on the
  source type. The first term corresponds to magnetic type sources
  while the second term corresponds to electric type sources.

  As can be seen, the above shows that the fields at two locations $\v
  r$ and $\v r'$ are correlated if they come from the same Langevin
  source.
  Most important, the above also shows that the two fields at two
  locations $\v r$ and $\v r'$ are correlated directly by the spectral
  function as in the left-hand side of \eqref{eqn91}, multiplied by
  $\Theta(\hbar \omega)$ and a multiplicative constant, which is a
  much simpler representation of the correlation function. The
  spectral function is non-singular at $\v r=\v r'$ and is seen to be
  set up by distributed Langevin sources.

  %Due to that the system is in thermal equilibrium, the spectral
  %function is described by a causal Green's function representing
  %field radiation into the environment by the Langevin sources, plus a
  %non-causal part representing field absorption from the environment
  %by the Langevin sources.
  %%
  % The lossless medium case as be
  %thought of as the limiting case of the above, as the Green's
  %function for a lossless medium is undefined unless infinitesimal
  %loss is introduced.

  \subsection{Energy Correlation Functions}

  In the above, we have derived the field correlation functions, and
  show their relationships to the spectral functions.
  To be more precise, when the energy density is needed,
  the energy correlation function needs to be derived as follows.
  An energy correlation function for the electric field, in connection
  to \eqref{eq:lrimc_5}, is then defined as
  \begin{align}
    \langle \dyadg\epsilon_e^{\frac12}(\v r)\cdot \v E(\v r,\omega)\v
    E^\dag(\v r',\omega')&\cdot\dyadg\epsilon_e^{\frac12}(\v
    r')\rangle\notag\\&=\delta(\omega-\omega')\dyadcal W_e(\v r,\v r',\omega)
  \end{align}
  In the above, $\dyadcal W_e(\v r,\v r',\omega)$ should have the
  dimension of energy density per radian frequency, and it can be
  connected to the electric spectral functions as
  \begin{align}
    \dyadcal W_e(\v r,\v r',\omega)=\frac{ \omega\mu_0}{\pi }
    \dyadg\epsilon_e^{\frac12}(\v r)\cdot \Im m\left[ \dyad G_e(\v r,\v
    r')\right]\cdot\dyadg\epsilon_e^{\frac12}(\v r') \Theta(\hbar
    \omega)
  \end{align}
  The above can be related to the LDOS when $\v r=\v r'$.

  Similarly, a correlation function for the magnetic field is defined
  as
  \begin{align}
    \langle \dyadg\mu_e^{\frac12 }(\v r)\cdot \v H(\v r,\omega)\v
    H^\dag(\v r',\omega')&\cdot\dyadg\mu_e^{\frac12}(\v
    r')\rangle\notag\\&=\delta(\omega-\omega')\dyadcal W_m(\v r,\v r',\omega)
  \end{align}
  With the understanding that $\dyadcal W_m(\v r,\v r',\omega)$ should
  have the dimension of energy density per radian frequency, it can be
  connected to the magnetic spectral functions as
  \begin{align}
    \dyadcal W_m(\v r,\v r',&\omega)\notag\\&=\frac{\omega\epsilon_0}{\pi}
    \dyadg\mu_e^{\frac12}(\v r)\cdot\Im m\left[\dyad G_m(\v r,\v
    r')\right]\cdot\dyadg\mu_e^{\frac12}(\v r') \Theta(\hbar \omega)
  \end{align}
  The above can be combined and are related to the LDOS defined in
  \eqref{eq:lrimc_6}.

  %\subsection{Relationship to FDT}
  %
  %
  %FDT says that if two quantities are linearly related, such as the
  %relation between the field and the random current via the Green's
  %function, a statement can be said about the field correlation. If
  %the current is randomly generated by thermal excitation, then the
  %correlation of the field is proportional to the imaginary part of
  %the Green's function \cite{Landau&Lifshitz,JoulainEtal}.
  %
  %Also, the Langevin currents are induced by random fields, and the
  %current is linearly related to the field by the permittivity or
  %permeability functions, then the correlation of the Langevin current
  %is proportional to the imaginary part of the permittivity or
  %permeability.
  %%
  %Furthermore, the exact form of the Langevin current correlation
  %functions can be derived by comparing the spectral function in
  %\eqref{eq:ldos74a} and \eqref{FDT:eq_91a}. We conclude that
  %\begin{align}\label{FDT:eq_93a}
  % \langle \v J(\v r, \omega)\v J^\dag (\v r',\omega')\rangle
  % =\delta(\omega-\omega')\delta(\v r-\v r')
  % \frac{\omega\epsilon}{\pi} \Im m\left[\dyadg \epsilon_r (\v
  % r)\right]\Theta(\hbar \omega)
  % \end{align}
  %and
  %\begin{align}\label{FDT:eq_94a}
  % \langle \v M(\v r, \omega)\v M^\dag (\v r',\omega')\rangle
  % =\delta(\omega-\omega')\delta(\v r-\v r')
  % \frac{\omega\mu}{\pi} \Im m\left[\dyadg \mu_r (\v
  % r)\right]\Theta(\hbar \omega)
  % \end{align}
  %Comparing \eqref{FDT:eq_93a} and \eqref{FDT:eq_94a}, one notes that
  %duality principle \cite{Kong} is satisfied.  The above can also be
  %derived by invoking the fluctuation dissipation theorem.

  \section{Conclusion}

  The Green's dyadics in terms of eigenmode expansions have been
  derived for general media including lossy, inhomogeneous,
  anisotropic media. Then the electromagnetic spectral functions are
  defined.  The spectral function as defined has the physical meaning
  of describing a source in equilibrium with its environment. It
  consists of a causal or radiative part, subtracting a non-causal, or
  absorptive part, representing the radiation and absorption of fields
  by Langevin sources.

  For lossless media, it can be shown that the spectral function is
  related to the local density of states.
  %
  %It also shows that the corresponding field mode amplitude is related
  %to the source frequency by a delta function. The sum of all possible
  %modes that are degenerate and associated with an operating frequency
  %$\omega$ yields the density of states or local density of states
  %associated with $\omega$.
  %
  For lossy media, due to line broadening, the spectral function
  describes a source at $\omega$ that excites a cluster of modes in
  time harmonic motion, even though these modes may not be degenerate.
  %One can assume that the summation over the cluster of modes yields
  %the density of states or local density of states at $\omega$. With
  %this assumption, the spectral function yields the same correlation
  %function for the field that is in thermal equilibrium with its
  %environment.
  Furthermore, by relating the spectral function to the field
  correlation function, the spectral function, including that for
  lossy anisotropic inhomogeneous media, can be related to the local
  density of states.

  Many relations derived using FDT can also be derived using spectral
  function.  One of them is that the correlation of the fields at two
  points is related via the spectral function which is the imaginary
  part of the Green's function. The second one is that the correlation
  of two Langevin currents are related via the imaginary part of their
  permittivity or permeability depending on the current type.
  Hence, our results are commensurate with FDT: namely, the same
  conclusions from spectral functions can be arrived at from FDT. It
  also justifies the use of FDT  for the calculation of Casimir force
  in a lossy medium.

\section*{Acknowledgements} This work was supported in part by the
  USA NSF CCF Award 1218552, SRC Award 2012-IN-2347, at the University
  of Illinois at Urbana-Champaign, by the Research Grants Council of
  Hong Kong (GRF 711609, 711508, and 711511), and by the University
  Grants Council of Hong Kong (Contract No. AoE/P-04/08) at HKU.
  The authors thank Carlos Salazar-Lazaro and Aiyin Liu for useful
  feedback.


\begin{thebibliography}{1}

\bibitem{Langevin}
P. Langevin, (1908). ``Sur la théorie du mouvement brownien [On the
Theory of Brownian Motion]". C. R. Acad. Sci. (Paris) 146: 530–533.

\bibitem{Johnson}
J. Johnson, ``Thermal Agitation of Electricity in Conductors", Phys.
Rev. 32, 97 (1928).

\bibitem{Nyquist}
 H. Nyquist, ``Thermal Agitation of Electric Charge in Conductors", Phys. Rev. 32, 110
 (1928).

 \bibitem{Haus}
 H.A. Haus. Electromagnetic noise and quantum optical measurements. Springer,
 2000.

\bibitem{Callen&Welton}
H.B. Callen, and T. A. Welton. ``Irreversibility and generalized
noise." Physical Review 83, no. 1 (1951): 34.


\bibitem{Rytov} S. Rytov, Y. Kravtsov, V. Tatarskii, Principles of Statistical Radiophysics, vol. 3, Springer-Verlag, Berlin,
1989.

%\bibitem{Kubo}
%Kubo, R. ``The fluctuation-dissipation theorem," Reports on Progress
%in Physics 29, no. 1 (1966): 255.

\bibitem{Kubo} R. Kubo, ``The fluctuation-dissipation theorem," Rep. Prog.
Phys., vol. 29, pp. 255-284, 1966.

\bibitem{Landau&Lifshitz}
L. D. Landau, E.M. Lifshitz, and L. P. Pitaevskii, Statistical
physics. Part 2. Oxford, UK: Pergamon Press, 1980.

\bibitem{Casimir}
H. B. G. Casimir, ``On the attraction between two perfectly
conducting plates," Proceedings of the Royal Netherlands Academy of
Arts and Sciences, vol. 51, pp. 793– 795, 1948.

\bibitem{Casimir&Polder} H. B. G. Casimir and D. Polder, ``The influence of retardation
on the London-van der Waals forces," Phys. Rev., vol. 73, pp.
360-372, 1948.


\bibitem{Lifshitz} E. M. Lifshitz, ``The theory of molecular attractive forces
between solids," Sov. Phys. JETP, vol. 2, p. 73, 1956.


\bibitem{VanKampenEtal} N. G. Van Kampen, B. R. A. Nijboer, and K. Schram, ``On
the macroscopic theory of van der Waals forces," Phys. Lett., vol.
26A, no. 7, pp. 307-308, 1968.



\bibitem{Milonni}
P.W. Milonni, The Quantum Vacuum: An Introduction to Quantum
Electrodynamics. San Diego, CA: Academic Press, 1994.

\bibitem{Lamoreaux} S. K. Lamoreaux, ``Demonstration of the casimir force in
the 0.6 to 6 $\mu$m range," Phys. Rev. Lett., vol. 78, no. 1, pp.
5-8, 1997.

\bibitem{Sernelius} B. E. Sernelius, ``Casimir force and complications
in the Van Kampen theory for dissipative systems," Phys. Rev. B,
vol. 74, p. 233103, 2006.

\bibitem{JohnsonSG}
A. Rodriguez, M. Ibanescu, D. Iannuzzi, J. D. Joannopoulos, and S.
G. Johnson, ``Virtual photons in imaginary time: Computing exact
Casimir forces via standard numerical electromagnetism techniques,"
Physical Review A, vol. 76, no. 3, p. 032106, 2007.

\bibitem{Capasso}
F. Capasso, J. Munday, D. Iannuzzi, and H. Chan, ``Casimir forces
and quantum electrodynamical torques: Physics and nanomechanics,"
IEEE Journal of Selected Topics in Quantum Electronics, vol. 13, no.
2, pp. 400–414, Mar.-Apr. 2007.

\bibitem{Capasso1} J. N. Munday, F. Capasso, and V. A. Parsegian, ``Measured
long-range repulsive Casimir-Lifshitz forces," Nature, vol. 457, p.
170, 2009.


\bibitem{RahiEtal} S. J. Rahi, T. Emig, N. Graham, R. L. Jaffe, and M. Kardar,
``Scattering theory approach to electrodynamic Casimir forces,"
Phys. Rev. D, vol. 80, p. 085021, 2009.

\bibitem{XiongEtal} J. L. Xiong, M. S. Tong, P. Atkins, and W. C. Chew, ``Efficient
evaluation of Casimir force in arbitrary three-dimensional
geometries by integral equation methods," Phys. Lett. A, vol. 374,
pp. 2517-2520, 2010.

\bibitem{RosaEtal} F. S. S. Rosa, D. A. R. Dalvit, and P. Milonni,
``Electromagnetic energy, absorption, and Casimir forces. ii.
inhomogeneous dielectric media," Phys. Rev. A, vol. 84, p. 053813,
2011.


\bibitem{Narayanaswamy&Chen} A. Narayanaswamy and G. Chen. ``Dyadic Green's functions and electromagnetic local density of
states." Journal of Quantitative Spectroscopy and Radiative Transfer
111, no. 12 (2010): 1877-1884.

\bibitem{GreffetEtal} J.-J. Greffet, R. Carminati, K. Joulain,
J.-P. Mulet, S. Mainguy, and Y. Chen, ``Coherent emission of light
by thermal sources." Nature 416, no. 6876 (2002): 61-64.

\bibitem{JoulainEtal} K. Joulain, J.-P. Mulet, F. Marquier, R. Carminati, J.-J.
Greffet,  ``Surface electromagnetic waves thermally excited:
Radiative heat transfer, coherence properties and Casimir forces
revisited in the near field." Surface Science Reports 57, no. 3
(2005): 59-112.

\bibitem{SprikEtal}
R. Sprik, B. A. van Tiggelen and A. Lagendijk, ``Optical emission in
periodic dielectrics," Europhys. Lett., 35 (4), pp. 265-270 (1996).

\bibitem{JonesEtal} A.C. Jones, B.T. O'Callahan, H.U. Yang, and M.B. Raschke
``The thermal near-field: Coherence, spectroscopy, heat-transfer,
and optical forces." Progress in Surface Science 88, no. 4 (2013):
349-392.

\bibitem{Keldysh}
L. V. Keldysh, L. V. ``Diagram technique for nonequilibrium
processes." Sov. Phys. JETP 20, no. 4 (1965): 1018-1026.

\bibitem{Datta}
S. Datta. Quantum transport: atom to transistor. Cambridge
University Press, 2005.

\bibitem{Tiggelen&Kogan}
B.A. Van Tiggelen and Eugene Kogan. ``Analogies between light and
electrons: Density of states and Friedel's identity." Physical
Review A 49, no. 2 (1994): 708.

\bibitem{Dirac} P.A.M. Dirac (1927). ``The Quantum Theory of the Emission and
Absorption of Radiation". Proceedings of the Royal Society of London
A 114 (767):
 243–265.

\bibitem{Glauber&Lewenstein}
R.J. Glauber and M. Lewenstein. ``Quantum optics of dielectric
media." Physical Review A 43, no. 1 (1991): 467.

\bibitem{Milonni1} P.W. Milonni, ``Field quantization and radiative processes in
dispersive dielectric media." Journal of Modern Optics 42, no. 10
(1995): 1991-2004.

\bibitem{Garrison&Chiao}
J.C. Garrison and R.Y. Chiao. Quantum optics. Oxford Univ. Press,
2008.


\bibitem{Hopfield} J.J. Hopfield. ``Theory of the contribution of excitons to the
complex dielectric constant of crystals." Physical Review 112, no. 5
(1958): 1555.

\bibitem{Caldeira&Leggett}
A. Caldeira and A. J. Leggett, ``Influence of dissipation on quantum
tunneling in macroscopic systems," Phys. Rev. Lett., vol. 46, p.
211, 1981.

\bibitem{Huttner&Barnett}
B. Huttner and S.M. Barnett. ``Quantization of the electromagnetic
field in dielectrics." Physical Review A 46, no. 7 (1992): 4306.

\bibitem{Grunner&Welsch} T. Grunner and D.-G. Welsch. ``Green-function approach to the
radiation-field quantization for homogeneous and inhomogeneous
Kramers-Kronig dielectrics." Physical Review A 53, no. 3 (1996):
1818.

\bibitem{DungEtal} H.T. Dung, L. Knoll, D.-G. Welsch. ``Three-dimensional quantization of the electromagnetic field in
dispersive and absorbing inhomogeneous dielectrics." Physical Review
A 57, no. 5 (1998): 3931.

\bibitem{DungEtal1} H.T. Dung, S.Y. Buhmann, L. Kn\"oll, D.-G.
Welsch, S. Scheel, and J. K\"astel. ``Electromagnetic-field
quantization and spontaneous decay in left-handed media." Physical
Review A 68, no. 4 (2003): 043816.

\bibitem{Scheel&Buhmann} S. Scheel and S.Y. Buhmann, ``Macroscopic Quantum Electrodynamics—Concepts And
Applications.'' Acta Physica Slovaca, vol. 58, No. 5, 675–809
October 2008.


\bibitem{Tai}  C.T. Tai. Dyadic Green functions in electromagnetic theory. Vol. 272. New
York: IEEE press, 1994.

\bibitem{Chew} W.C. Chew.
Waves and fields in inhomogeneous media. Vol. 522. New York: IEEE
press, 1995 (First Printing, Van Nostrand Reinhold, 1990).

\bibitem{ChewTongHu}
W. C. Chew, M.S. Tong, and B. Hu, {\it Integral equation methods for
electromagnetic and elastic waves.} Morgan \& Claypool Publishers,
2008.


\bibitem{Gerry&Knight}
C. Gerry and P. Knight. Introductory quantum optics. Cambridge
university press, 2005.

\bibitem{Fox} M. Fox. Quantum Optics: An Introduction. Vol. 6. Oxford
University Press, 2006.

\bibitem{Feynman}
R.P. Feynman, R.B. Leighton, and M. Sands. {\it The Feynman Lectures
on Physics}, Vol. 1, 41-1, Basic Books, 2013.

\bibitem{Kramers&Kronig} R. Kronig and H. A. Kramers. "Absorption and dispersion in
x-ray spectra." Z. Phys 48 (1928): 174.


\bibitem{DaiEtal} Q.I. Dai, Y.H. Lo, W.C. Chew, Y.G. Liu, and L.J. Jiang,
``Generalized modal expansion and reduced modal representation of
3-D electromagnetic fields," IEEE Transactions on Antennas and
Propagation, vol. 62, pp. 783-793, Feb 2014.

\bibitem{Jones&Raschke} A.C. Jones, M. Raschke. ``Thermal infrared
near-field spectroscopy." Nano letters 12, no. 3 (2012): 1475-1481.

\bibitem{SapienzaEtal} R. Sapienza, T. Coenen, J. Renger, M. Kuttge, N.F. van
Hulst, and A. Polman. ``Deep-subwavelength imaging of the modal
dispersion of light." Nature materials 11, no. 9 (2012): 781-787.

\bibitem{Loudon} R. Loudon. ``The propagation of electromagnetic energy through an
absorbing dielectric." Journal of Physics A: General Physics 3, no.
3 (1970): 233.

\bibitem{Ruppin} R. Ruppin, ``Electromagnetic energy density in a dispersive and
absorptive material." Physics letters A 299, no. 2 (2002): 309-312.


%\bibitem{Economou} "Green's Functions in Quantum Physics, Eleftherios N.
%Economou, 1978, Springer".

\bibitem{Busch&John} K. Busch and S. John. ``Photonic band gap formation
in certain self-organizing systems." Physical Review E 58, no. 3
(1998): 3896.



\bibitem{Kong}
J.~A. Kong, {\it Theory of electromagnetic waves}. New York,
Wiley-Interscience, 1975. 348 p. 1 (1975).













\end{thebibliography}
\end{document}